\documentclass[format=acmsmall, review=false]{acmart}
\usepackage{acm-ec-22}
\usepackage{booktabs} % For formal tables
\usepackage[ruled]{algorithm2e} % For algorithms

\SetAlFnt{\small}
\SetAlCapFnt{\small}
\SetAlCapNameFnt{\small}
\SetAlCapHSkip{0pt}
\IncMargin{-\parindent}

\usepackage[hyphenbreaks]{breakurl}

\newtheorem{example}{Example}
\newtheorem{theorem}{Theorem}
\newtheorem{lemma}{Lemma}
\newtheorem{conjecture}{Conjecture}

\newtheorem{fact}{Fact}
\newtheorem{definition}{Definition}

\newtheorem{proposition}{Proposition}

\newcommand{\R}{\mathbb{R}} % Real numbers
\newcommand{\N}{\mathbb{N}} % Integer numbers
\newcommand{\ind}[1]{\mathbf{1}\left\{ #1\right\}} % Indicator

\newcommand{\E}{\mathbb{E}} % Expected value
\newcommand{\bP}{\mathbb{P}} % Probability

\newcommand{\subjectto}{\text{\rm subject to}} % subject to
\newcommand{\sN}{\mathcal{N}} % Agents
\renewcommand{\G}{\mathcal{G}} % Groups
\newcommand{\X}{\mathcal{X}} % Deterministic allocations
\newcommand{\AllocationSpace}{\Delta(\X)} % Lottery allocations
\newcommand{\B}{\mathcal{B}} % Set of strategies used in our conjecture
\newcommand{\s}{\mathcal{S}}
\newcommand{\positions}{P}
\newcommand{\ra}{\pi} %
\newcommand{\xil}{x^{IL}}
\newcommand{\xspl}{x^{IW}}
\newcommand{\xgl}{x^{GL}}

\newcommand{\IL}{\pi^{IL}}
\newcommand{\SPL}{\pi^{IW}}
\newcommand{\GL}{\pi^{GL}}
\newcommand{\GLR}{\pi^{GR}}
\renewcommand{\a}{{\bf a}}
\newcommand{\order}{{\sigma}}
\newcommand{\rOrder}{{\Sigma}}
\newcommand{\sq}{\Sigma}
\renewcommand{\o}{\mathcal{O}}
\newcommand{\NameProposedMechanism}{{Weighted Individual Lottery}}

\setcitestyle{authoryear}

\title[Lotteries for Shared Experiences]{Lotteries for Shared Experiences}
\subtitle{May 2022}
\author{Nick Arnosti}
\affiliation{%
  \institution{University of Minnesota}
  \city{Minneapolis}
  \state{MN}
  \country{US}
}
\email{arnosti@umn.edu}

\author{Carlos Bonet}
\affiliation{%
  \institution{Columbia University}
  \city{New York}
  \state{NY}
  \country{US}
}
\email{cbonet23@gsb.columbia.edu}
\begin{abstract}
We study a setting where tickets for an experience are allocated by lottery. Each agent belongs to a group, and a group is {\em successful} if and only if its members receive enough tickets for everyone. A lottery is {\em efficient} if it maximizes the number of agents in successful groups, and {\em fair} if it gives every group the same chance of success. We study the efficiency and fairness of existing approaches, and propose practical alternatives.

If agents must identify the members of their group, a natural solution is the {\em Group Lottery}, which orders groups uniformly at random and processes them sequentially. We provide tight bounds on the inefficiency and unfairness of this mechanism, and describe modifications that obtain a fairer allocation.

If agents may request multiple tickets without identifying members of their group, the most common mechanism is the {\em Individual Lottery}, which orders agents uniformly at random and awards each their request until no tickets remain. Because each member of a group may apply for (and win) tickets, this approach can yield arbitrarily unfair and inefficient outcomes. As an alternative, we propose the {\em {\NameProposedMechanism}}, in which the processing order is biased against agents with large requests. Although it is still possible to have multiple winners in a group, this simple modification makes this event much less likely. As a result, the {\NameProposedMechanism} is approximately fair and approximately efficient, and similar to the Group Lottery when there are many more agents than tickets.
\end{abstract}

\begin{document}

% Title page for title and abstract only.
\begin{titlepage}
\maketitle
\end{titlepage}

% Paper body
\section{Introduction} \vspace{.1 in}

\subsection{Motivation}

Although matching models often assume that agents care only about their own allocation, there are many scenarios where people also care about the allocation received by their friends or family members. For example, couples entering residency may wish to be matched to programs in the same region, siblings may wish to attend the same school, and friends may want to share a hiking trip. Practitioners often employ ad-hoc solutions in an effort to accommodate these preferences.

This paper studies a special case of this problem, in which there are multiple copies of a homogeneous good. Each agent belongs to a group, and is successful if and only if members of her group receive enough copies for everyone in the group.
Examples of such settings include:
\begin{itemize}
\item {\em American Diversity Visa Lottery.} Each year 55,000 visas are awarded to citizens of eligible countries. Applicants are selected by lottery. Recognizing that families want to stay together, the state department grants visas to eligible family members of selected applicants.\footnote{Details of the 2022 Diversity Immigrant Visa Program are available at \url{https://travel.state.gov/content/dam/visas/Diversity-Visa/DV-Instructions-Translations/DV-2022-Instructions-Translations/DV-2022-Instructions-and-FAQs_English.pdf}.}

\item {\em Big Sur Marathon.} Many popular marathons limit the number of entrants and use a lottery to select applicants. The Big Sur Marathon uses several lotteries for different populations (i.e. locals, first-timers, and returning runners from previous years). One of these is a ``Groups and Couples" lottery which ``is open for groups of from 2-15 individuals, each of whom want to run the Big Sur Marathon but only if everyone in the group is chosen." In 2020, 702 tickets were claimed by 236 successful groups selected from 1296 applicants.
\footnote{More information about the 2020 Big Sur Marathon Drawing is available at \url{https://www.bigsurmarathon.org/random-drawing-results-for-the-2020-big-sur-marathon/}}

\item {\em Hiking Permits on Recreation.gov.} Many parks use a permit system to limit the number of hikers on popular trails. For example, the permits to hike Half Dome in Yosemite National Park are awarded through a pre-season lottery, as well as daily lotteries.\footnote{More information available at \url{https://www.nps.gov/yose/planyourvisit/hdpermits.htm}} To enable applicants to hike with friends and family, each applicant is allowed to apply for up to six permits.

\item {\em Discounted Broadway Tickets.} Many popular Broadway shows hold lotteries for discounted tickets. While some people may be happy going to a Broadway show alone, most prefer to share the experience with others. Recognizing this fact, theaters typically allow each applicant to request up to two tickets. On the morning of the show, winners are selected and given the opportunity to purchase the number of tickets that they requested.

%\todo{\item Add an example where the GL is being implemented.}
\end{itemize}
Inspired by the last application, in the rest of the paper we will refer to a copy of the homogeneous good as a `ticket'.

The settings above present several challenges. First and foremost, the designer must prevent individuals from submitting multiple applications. In high-stakes environments such as the diversity visa lottery, this can be accomplished by asking applicants to provide government identification as part of their application. In applications with lower stakes, this is frequently accomplished by tying each application to an e-mail address, phone number, or social media account. The effectiveness of this approach will vary across settings. If the designer is concerned that individuals may be submitting multiple applications, then this concern should be addressed before anything else. In this paper, we assume that the designer has a way to identify each individual, and verify that nobody has submitted duplicate applications.

A second challenge is that designers do not know who belongs to each group. One solution is to ask applicants to identify members of their group in advance. While this is done for the diversity visa lottery and for affordable housing lotteries, it can be quite cumbersome. It requires additional effort from applicants, which may be wasted if their applications are not selected. In addition, to ensure that applicants do not submit false names, when awarding tickets the designer must verify that the identity of each recipient matches the information on the application form. Perhaps for these reasons, many designers opt for a simpler interface which allows applicants to specify how many tickets they wish to receive, but does not ask them to name who these tickets are for.

Motivated by these observations, we study two types of mechanisms:  ``direct" mechanisms which ask applicants to identify members of their group, and mechanisms which only ask each applicant to specify a number of tickets requested. In the former case, the most natural approach is to place groups in a uniformly random order, and sequentially allocate tickets until no more remain. This procedure, which we refer to as the {\em Group Lottery}, is used, for example, to allocate affordable housing in New York City. In the latter case, an analogous procedure is often used: applicants are processed in a uniformly random order, with each applicant given the number of tickets that they requested until no tickets remain. We call this mechanism the {\em Individual Lottery}, and variants of it are used in all of the applications listed above.\footnote{
Recreation.gov goes into great detail about the algorithm used to generate a uniform random order of applicants (\url{https://www.recreation.gov/lottery/how-they-work}), while the FAQ for the Diversity Visa Lottery notes, ``a married couple may each submit a DV Lottery application and if either is selected, the other would also be entitled to a permanent resident card" (\url{https://www.dv-lottery.us/faq/}).}

\subsection{Concerns with existing approaches}

Although the Individual Lottery and the Group Lottery seem natural and are used in practice, they each have flaws. In the Individual Lottery, each member of a group can submit a separate application. This is arguably arguably {\em unfair}, as members of large groups might have a much higher chance of success than individual applicants. In addition, the Individual Lottery may be {\em inefficient}. One reason for this is that there is no penalty for submitting a large request, so some individuals may ask for more tickets than their group needs.\footnote{Applicants are very aware of this. One of the authors received an e-mail from the organizer of a Half Dome trip who noted, ``It costs nothing extra to apply for 6 spots. If you do win, you might as well win big!" Meanwhile, a guide about the lottery for the Broadway show Hamilton advises, ``You can enter the lottery for either one or two seats. Always enter it for two. A friend you bring to Hamilton will be a friend for life" (\url{https://www.timeout.com/newyork/theater/hamilton-lottery}).} Even if this does not occur, multiple members of a group might apply and win tickets, resulting in some of these tickets going to waste.

Anecdotally, we see strong evidence of groups with multiple winners in the Big Sur Marathon lottery. Although the information page suggests that ``a single, designated group leader enters the drawing on behalf of the group," in 2019, the lottery winners included two teams titled ``Taylor's" (with leaders Molly Taylor and Amber Taylor, respectively), as well as a team titled ``What the Hill?" and another titled ``What the Hill?!"\footnote{More information about the drawing can be found at \url{https://web.archive.org/web/20200407192601/https://www.bigsurmarathon.org/drawing-info/}. The list of groups awarded in 2019 is available at \url{http://www.bigsurmarathon.org/wp-content/uploads/2018/07/Group-Winners-for-Website.pdf}}. These examples suggest that groups are (rationally) not abiding by the recommendation that only one member enter the lottery, and that some groups are receiving more tickets than needed.

The instruction that only one member of each group should apply to the Big Sur Marathon lottery suggests that the organizers intended to implement a Group Lottery. Although the Group Lottery overcomes some of the issues described above, it is also not perfectly fair nor perfectly efficient. This is because when only a few tickets remain, (i) small groups still have a chance of success while large groups do not, and (ii) these tickets may be wasted if the next group to be processed is large. Because these issues arise only at the end of the allocation process, one might hope that the resulting allocation will not be too unfair or inefficient.

Our first contribution is to quantify the unfairness and inefficiency of these mechanisms. While the fairness and efficiency of the Individual Lottery suffer when everyone from a group applies, how bad can the problem be? And is the intuition that the Group Lottery is approximately fair and efficient correct? Our answers to these questions are ``very bad", and ``yes," respectively. Although neither mechanism is perfectly fair or efficient, there is a large qualitative and quantitative difference between them. Our second contribution is to identify modifications to each algorithm which use the same user interfaces but offer improved fairness and/or efficiency. We elaborate on these contributions below.

\subsection{Overview of Model and Results}

We consider a model with $k$ identical tickets. The set of agents is partitioned into a set of groups, and agents have {\em dichotomous preferences}: an agent is successful if and only if members of her group receive enough tickets for everyone in the group. We treat the group structure as private information, unknown to the designer. Because there are only $k$ tickets, there can be at most $k$ successful agents. We define the efficiency of a lottery allocation to be the expected number of successful agents, divided by $k$. If this is at least $\beta$, then the allocation is {\em $\beta$-efficient}. A lottery allocation is {\em fair} if each agent has the same success probability, and {\em $\beta$-fair} if for any pair of agents, the ratio of their success probabilities is at least $\beta$.

Given these definitions, we seek lottery allocations that are both approximately efficient and approximately fair. Although this may be unattainable if groups are large, in many cases group sizes are much smaller than the total number of tickets. We define a family of instances characterized by two parameters, $\kappa$ and $\alpha$. The parameter $\kappa$ bounds the ratio of group size to total number of tickets, while $\alpha$ bounds the supply-demand ratio. For any $\kappa$ and $\alpha$, we provide worst-case performance guarantees in terms of efficiency and fairness.

We first consider a scenario where applicants can identify each member of their group. Here, the mechanism typically used is the Group Lottery. We show in Proposition~\ref{prop:GL-incentives} that this mechanism incentivizes agents to truthfully report their groups. Moreover, Theorem~\ref{thm:gl-is-good} establishes that the Group Lottery is $(1 - \kappa)$-efficient and \((1-2\kappa)\)-fair. It is not perfectly efficient, as tickets might be wasted if the size of the group being processed exceeds the number of remaining tickets. It is not perfectly fair, since once only a few tickets remain, a large group can no longer be successful, but a small group can. Proposition \ref{prop:gl-tightness} shows that this guarantee is tight.

Could there be a mechanism with stronger performance guarantees than the Group Lottery? Proposition~\ref{prop:badnews} establishes the limits of what can be achieved. Specifically, it says that there always exists an allocation \(\pi\) that is $(1-\kappa)$-efficient and fair, but for any $\epsilon > 0$, there are examples where any allocation that is $(1- \kappa + \epsilon)$-efficient is not even $\epsilon$-fair. To show the existence of the random allocation \(\pi\), % that achieves these guarantees,
we use a generalization of the Birkhoff-von Neumann theorem proved by \citet{nguyen2016assignment}. By awarding groups according to the allocation \(\pi\), we can obtain a mechanism that attains the best possible performance guarantees. Therefore, the $2 \kappa$ loss in fairness in the Group Lottery can be thought of as the ``cost" of using a simple procedure that orders groups uniformly, rather than employing a Birkhoff-von Neumann decomposition to generate the allocation \(\pi\).

In many applications, developing an interface that allows applicants to list their group members may be too cumbersome. This motivates the study of a second scenario, where applicants are only allowed to specify the number of tickets they need. The natural mechanism in this setting is the Individual Lottery. Unfortunately, Theorem~\ref{thm:il-is-bad} establishes that the Individual Lottery may lead to arbitrarily inefficient and unfair outcomes. It is perhaps not surprising that the Individual Lottery will be inefficient if agents request more tickets than needed, or if each agent has a large chance of success. However, we show that the waste due to over-allocation may be severe even if all agents request only their group size and demand far exceeds supply. Furthermore, because the probability of success will be roughly proportional to group size, small groups are at a significant disadvantage.

Can we achieve approximate efficiency and fairness without asking applicants to identify each member of their group? We show that this is possible with a minor modification to the Individual Lottery which gives applicants with larger requests a lower chance of being allocated. This eliminates the incentive to inflate demand, and reduces the possibility of multiple winners from the same group. To make the allocation fair, we choose a particular method for biasing the lottery against large requests: sequentially select individuals with probability inversely proportional to their request. We call this approach the {\em \NameProposedMechanism}. In the {\NameProposedMechanism}, a group of four individuals who each request four tickets has the same chance of being drawn next as a group of two individuals who each request two tickets. As a result, outcomes are similar to the Group Lottery.
We prove that the {\NameProposedMechanism} is (\(1-\kappa-\alpha/2\))-efficient and (\(1-2\kappa-\alpha/2\))-fair (in fact, Theorem~\ref{thm:spl-performance} establishes slightly stronger guarantees). Notice that these guarantees  coincide with those of the Group Lottery when demand far exceeds supply ($\alpha$ is close to $0$).

Our main results are summarized in Table ~\ref{tab:main-results}. Our conclusion is that the Individual Lottery can be arbitrarily unfair and inefficient. These deficiencies can be mostly eliminated by using a Group Lottery. Perhaps more surprisingly, approximate efficiency and fairness can also be achieved while maintaining the Individual Lottery interface, by suitably biasing the lottery against agents with large requests.

\begin{table}[t]
\centering
\begin{tabular}{l c l l}
 \hline
Mechanism & Action Set  \hspace{.2 in}  & Efficiency \hspace{.1 in} & Fairness \\ \hline
Benchmark & & $1 - \kappa$ & 1\\
Group Lottery & $2^\sN$ & $1 - \kappa$ &  $1 - 2\kappa$ \\
Individual Lottery  & $\{1, 2, \ldots, k\}$  & 0 & 0\\
Weighted Individual Lottery & $\{1, 2, \ldots, k\}$ & $1 - \kappa -\alpha/2$ & \(1-2\kappa-\alpha/2\) \\\hline

\end{tabular}
\caption{\label{tab:main-results}Summary of main results: worst-case guarantees for the efficiency and fairness of instances in \(I(\kappa, \alpha)\). These guarantees are established in Theorems~\ref{thm:gl-is-good},~\ref{thm:il-is-bad} and~\ref{thm:spl-performance}. Meanwhile, Proposition \ref{prop:badnews} establishes that the best one can hope for is a mechanism that is $(1 - \kappa)$ efficient and $1$-fair. }
\end{table}

\section{Related work}

Our high-level goal of allocating objects efficiently, subject to fairness and incentive compatibility constraints, is shared by numerous papers. The definitions of efficiency, fairness, and incentive compatibility differ significantly across settings, and below, we focus on papers that are closely related to our own.

If the group structure is known to the designer, then our problem simplifies to allocating copies of a homogeneous item to groups with multi-unit demand. This problem has received significant attention. \citet{benassy_1982} introduces the uniform allocation rule, in which each group requests a number of copies, and receives the minimum of its request and a quantity $q$, which is chosen so that every copy is allocated. \citet{sprumont_1991} and \citet{ching_1992} show that when preferences are single-peaked, this is the unique rule that is Pareto efficient, envy-free, and incentive compatible. \citet{ehlers2003probabilistic} extend this characterization to randomized allocation mechanisms. \citet{cachon1999capacity} consider the uniform allocation rule in a setting where groups have decreasing marginal returns from additional items.

In contrast to these papers, we assume that groups have dichotomous preferences, with no value for receiving only a fraction of their request. As a result, uniform allocation would be extremely inefficient. Instead, we propose the Group Lottery, which resembles the ``lexicographic" allocation rule from \citet{cachon1999capacity}. Dichotomous preferences have also been used to model preferences in kidney exchange \cite{roth2005pairwise}, two-sided matching markets \cite{bogomolnaia2004random}, and collective choice problems \cite{bogomolnaia2005collective}.

The all-or-nothing nature of preferences means that our work is related to the ``fair knapsack" problem introduced by \citet{patel2020group}, where a planner must choose a subset of groups to allocate, subject to a resource constraint. Groups are placed into categories, and the number of successful groups from each category must fall into specified ranges. Their model is fairly general, and if groups are categorized by size, then ranges can be chosen to make their fairness notion similar to ours. However, they do not quantify the cost of imposing fairness constraints. By contrast, we show that in our setting, fairness can be imposed with little or no cost to efficiency. Furthermore, approximate efficiency and fairness can be achieved in our setting using mechanisms that are much simpler than their dynamic-programming based algorithms.

Closer to our work is that of \citet{nguyen2016assignment}. They consider a setting in which each group has complex preferences over bundles of heterogeneous items, but only wants a small fraction of the total number of items. They find approximately efficient and fair allocations using a generalization of the Birkhoff-von Neumann theorem. Although their notion of fairness is different from ours, we use their results to prove Proposition~\ref{prop:badnews}. However, our papers have different goals: their work identifies near-optimal but complex allocation rules, while we study the performance of simple mechanisms deployed in practice, and close variants of these mechanisms.

An important difference between all of the aforementioned papers and our own is that we assume that the group structure is unknown to the designer. In theory, this can be solved by asking agents to identify the members of their group (as in the Group Lottery), but in many contexts this may be impractical. Hence, much of our analysis considers a scenario where agents are asked to report only a single integer (interpreted as the size of their group). We show what can be achieved is this setting, through our analysis of the Individual Lottery and {\NameProposedMechanism}. We are unaware of any prior work with related results.

We close by highlighting two papers with results that are used in our analysis. \citet{serfling1974probability} introduces the martingale when sampling without replacement from a finite population. This martingale is key in the proof of Proposition~\ref{lemma:mart-bounds}, which establishes bounds on the expected hitting time for the sample sum. This, in turn, is used to establish our fairness result for the Group Lottery. \citet{johnson2005univariate} state a simple bound on the probability that a Poisson random variable deviates from its expectation at least by a given number. We use this result in our analysis of the {\NameProposedMechanism}, where we use a Poisson random variable to bound the probability that a group has at least \(r\) members awarded.

\section{The model}

\subsection{Agents, Outcomes, Utilities}
A designer must allocate $k\in\N$ indivisible identical tickets to a set of agents $\sN = \{1,...,n\}$. A \emph{feasible allocation} is represented by $x \in \{0,1,...,k\}^n$ satisfying $\sum_{i\in \sN}x_i \le k$, where $x_i$ indicates the number of tickets that agent $i$ receives. We let $\X$ be the set of all feasible allocations.

A \textit{lottery allocation} is a probability distribution $\ra$ over $\X$, with $\ra_x$ denoting the probability of allocation $x$. Let $\AllocationSpace$ be the set of all lottery allocations.

The set $\sN$ is partitioned into groups according to $\G$: that is, each $G \in \G$ is a subset of $\sN$, $\cup_{G\in\G}G = \sN$, and for each $G,G'\in\G$ either $G = G'$ or $G\cap G' = \emptyset$. Given agent $i\in \sN$, we let $G_i\in\G$ be the group containing $i$. Agents are successful if and only if the total number of tickets allocated to the members of their group is at least its cardinality. Formally, each $i\in \sN$ is endowed with a utility function $u_i: \X \to \{0,1\}$ given by
\begin{equation} u_{i}(x) = \ind{\sum_{j\in G_i} x_{j}\ge |G_i|}.\end{equation}
We say that agent $i$ is {\em successful} under allocation $x$ if $u_i(x) = 1$.

In a slight abuse of notation, we denote the expected utility of agent $i\in \sN$ under the lottery allocation $\ra$ by
\begin{equation} u_i(\ra) = \sum_{x \in \X} \ra_x u_i(x).\end{equation}

\subsection{Performance Criteria}

We define the \textit{expected utilization} of a lottery allocation $\ra$ to be
\begin{equation}\label{def:lot_utilization}
U(\ra) = \frac{1}{k}\sum_{i\in \sN}u_i(\ra).
\end{equation}
\begin{definition}[Efficiency]
{\em A lottery allocation $\ra$ is {\bf efficient} if $U(\ra) = 1.$ It is {\bf $\beta$-efficient} if $U(\ra)\ge \beta$.}
\end{definition}

\begin{definition}[Fairness]
{\it A lottery allocation $\ra$ is  {\bf fair} if for every $i,i' \in \sN,$ $u_i(\ra) = u_{i'}(\ra)$. It is {\bf $\beta$-fair} if for every $i, i' \in \sN$, $u_i(\ra) \ge \beta u_{i'}(\ra)$.}
\end{definition}

\subsubsection*{Alternative Definitions of Fairness.}

With dichotomous preferences, our notion of efficiency seems quite natural. Our fairness definition states that agents in groups of different sizes should have similar expected utilities. There are other notions of fairness that one might consider. Two that arise in other contexts are {\em equal treatment of equals} and {\em  envy-free}. Below we present the natural analog of these in our setting, and discuss their relation to our definition.
\begin{definition}
{\it Lottery allocation $\ra$ satisfies {\bf equal treatment of equals} if for every pair of agents \(i,j\) such that \(|G_i| = |G_j|\), we have $u_i(\ra) = u_j(\ra)$.}
\end{definition}
This is clearly weaker than our fairness definition, and is an easy property to satisfy. In particular, the group request outcomes of the three mechanisms we study all satisfy equal treatment of equals.

To define envy-freeness, we introduce additional notation. For any \(x \in \X\), we let $N_G(x)$ be the number of tickets allocated to members of $G$. For any \(\ra \in \AllocationSpace\), let $N_G(\ra)$ be a random variable representing this number. Let $u_G(N) = \bP(N \geq |G|)$ be the expected utility of $G$ when the number of tickets received by members of $G$ is equal to $N$.

\begin{definition}
{\it Lottery allocation $\ra$ is {\bf group envy-free} if no group envies the allocation of another: $u_G(N_G(\ra)) \geq u_{G}(N_{G'}(\ra))$ for all $G, G' \in \G$.}
\end{definition}

This notion is neither stronger nor weaker than our fairness definition. To see this, suppose that there is a group of size $1$ and another of size $2$. The group of size $1$ gets one ticket with probability $\epsilon$, and otherwise gets zero tickets. The group of size $2$ gets two tickets with probability $\epsilon$ and otherwise gets one ticket. This is fair (according to our definition) but not even approximately group envy-free. Conversely, if both groups get one ticket with probability $\epsilon$ and zero tickets otherwise, then the allocation is group envy-free but not fair.

However, the conclusions we draw would also hold for this new fairness notion. The group request outcome of the Individual Lottery may not be even approximately group envy-free. The group request outcome of the Group Lottery is group envy-free. The group request outcome of the {\NameProposedMechanism} is approximately group envy-free, with the same approximation factor as in Theorem~\ref{thm:spl-performance}.

\subsection{Actions and Equilibria}

The designer can identify each agent (and therefore prevent agents from applying multiple times), but does not know the group structure a priori. Therefore, the designer must deploy a mechanism that asks individual agents to take actions. When studying incentives induced by a mechanism, however, we assume that members of a group can coordinate their actions.

Formally an anonymous {\em mechanism} consists of an action set $A$ and an allocation function $\pi : A^{\sN} \rightarrow \AllocationSpace$, which specifies a lottery allocation $\pi(\a)$ for each possible {\em action profile} $\a \in A^{\sN}$.

\begin{definition}\label{def:dominant-strategy}
{\it The actions $\a_{G_i} \in A^{G_i}$ are {\bf dominant} for group $G_i$ if for any actions $\a_{-G_i} \in A^{\sN \backslash G_i}$,
\begin{equation}\a_{G_i} \in \arg\max_{\a'_{G_i}\in A^{G_i}} u_i(\pi(\a'_{G_i},\a_{-G_i})).\label{eq:opt}\end{equation}
The action profile $\a \in A^{\sN}$ is a {\bf dominant strategy equilibrium} if for each group $G\in\G$, actions $\a_{G}$ are dominant for $G$.}
\end{definition}

Note that although actions are taken by individual agents, our definition of dominant strategies allows all group members to simultaneously modify their actions. We believe that this reasonably captures many settings, where group members can coordinate but the mechanism designer has no way to identify groups a priori.

\section{Results}

In general, it might not be feasible to achieve approximate efficiency, even if the group structure is known. Finding an efficient allocation involves solving a knapsack problem where each item represents a group, and the knapsack's capacity is the total number of tickets. If it is not possible to select a set of groups whose sizes sum up to the number of tickets, then some tickets will always be wasted. This issue can be particularly severe if groups' sizes are large relative to the total number of tickets. Therefore, one important statistic will be the ratio of the maximum group size to the total number of tickets.

Additionally, one concern with the Individual Lottery is that tickets might be wasted if groups have multiple winners. Intuitively, this is more likely when the number of tickets is close to the number of agents. Therefore, a second important statistic will be the ratio of tickets to agents.

These thoughts motivate us to define a family of instances characterized by two parameters: \(I(\kappa,\alpha)\). The parameter $\kappa$ captures the significance of the ``knapsack" structure, and $\alpha$ captures the ``abundance" of the good. For any \(\kappa,\alpha \in (0,1)\), define the family of instances \(\)
\begin{equation}\label{alpha-kappa-family}
I(\kappa, \alpha) = \left\{(n,k,\G) :\frac{\max_{G\in\G} |G|-1}{k} \leq \kappa, \frac{k}{n} \leq \alpha \right\}.
\end{equation}

Therefore, when analyzing a mechanism we study the worst-case efficiency and fairness guarantees in terms of $\kappa$ and $\alpha$. Ideally, we might hope for a solution that is both approximately fair and approximately efficient. Theorem~\ref{thm:gl-is-good} shows that this is achieved by the Group Lottery, which asks agents to reveal their groups. By contrast, Theorem~\ref{thm:il-is-bad} establishes that the Individual Lottery may lead to arbitrarily inefficient and unfair outcomes. Finally, Theorem~\ref{thm:spl-performance} establishes that the {\NameProposedMechanism} is approximately fair and approximately efficient, and similar to the Group Lottery when there are many more agents than tickets ($\alpha$ is small).

\subsection{Group Lottery}\label{S:GL}

In this section, we present the Group Lottery ($GL$) and show that it is approximately fair and approximately efficient. In this mechanism, each agent is asked to report a subset of agents, interpreted as their group. We say that a group $S \subseteq \sN$ is valid if all its members declared the group $S$. Valid groups are placed in a uniformly random order and processed sequentially (agents that are not part of a valid group will not receive tickets). When a group is processed, if enough tickets remain then every member of the group is given one ticket. Otherwise, members of the group receive no tickets and the lottery ends.\footnote{There is a natural variant of this mechanism which skips over large groups when few tickets remain, and gives these tickets to the next group whose request can be accommodated. This variant may be arbitrarily unfair, as can be seen by considering an example with an odd number of tickets, one individual applicant, and many couples. Then the individual applicant is always successful, while the success rate of couples can be made arbitrarily small by increasing the number of couples.}

We now introduce notation that allows us to study this mechanism. For any finite set \(E\), we let \(\s_E\) be the set of finite sequences of elements of \(E\), and let \(\o_{E}\) be the set of sequences such that each element of \(E\) appears exactly once. We refer to an element \(\sigma \in \o_{E}\) as an {\em order} over \(E\), with \(\sigma_t \in E\) and \(\sigma_{[t]} = \bigcup_{t' \leq t} \sigma_{t'}\) denoting the subset of \(E\) that appears in the first \(t\) positions of \(\sigma\).

Next we provide a formal description of the mechanism. The action set is the power set of $\sN$. Given an action profile $\a$, we call a set of agents $S \subseteq \sN$ a \textit{valid group} if for every agent $i \in S$ we have that $a_i = S$. We define a function \(\tau\) that will let us to characterize the number of valid groups that obtain their full request.
Fix a finite set \(E\) and a size function $| \cdot | : E \rightarrow \N$. For any $c \in \N$ and $\sigma \in \s_E$ satisfying $\sum_{t} |\sigma_t| \geq c$, define
\begin{equation}
\tau(c,\order) = \min\left\{T \in \N: \sum_{t=1}^T |\order_t| \ge c\right\}. \label{eq:def:tau}
\end{equation}

Fix an arbitrary action profile \(\a\) and let $V$ be the resulting set of valid groups. For any order $\order \in \o_V$, we let $\tau = \tau(k+1, \order)$ be as in \eqref{eq:def:tau} where the size of each valid group is its cardinality. Then the number of valid groups that are processed and obtain their full request is \(\tau - 1\). We define
\begin{equation}\label{eq:xgl}
\xgl_i(\a,\sigma) = \sum_{j=1}^{\tau-1} \ind{i \in \sigma_j}.
\end{equation}
For any \(x' \in \X\), the allocation function of the Group Lottery is
\[\GL_{x'}(\a) = \sum_{\order \in \o_V}\ind{x' = \xgl(\a,\order)}\frac{1}{|V|!}.\]

\subsubsection{Incentives.}

In every mechanism that we study, there is one strategy that intuitively corresponds to truthful behavior. We refer to this as the \emph{group request strategy}. In the Group Lottery, this is the strategy in which each agent declares the members his or her group.

\begin{definition}\label{def:GL-GR}
{\em In the Group Lottery, group $G \in \G$ follows the  {\bf group request} strategy if $a_i = G$ for all $i \in G$.}
\end{definition}

\begin{proposition}\label{prop:GL-incentives}
In the Group Lottery, the group request strategy is the only dominant strategy.
\end{proposition}
The intuition behind Proposition \ref{prop:GL-incentives} is as follows. Potential deviations for group $G$ include splitting into two or more groups, or naming somebody outside of the group as a member. We argue that in both cases the group request is weakly better. First, neither approach will decrease the number of other valid groups. Second, if there are at least $|G|$ tickets remaining and a valid group containing a member of $G$ is processed, then under the group request $G$ gets a payoff of 1. This might not be true under the alternatives strategies.

In light of Proposition \ref{prop:GL-incentives}, we will assume that groups follow the group request strategy when analyzing the performance of the Group Lottery.

\subsubsection{Performance.}

We next argue that the Group Lottery is approximately fair and approximately efficient.
Of course, the Group Lottery is not perfectly efficient, as it solves a packing problem greedily, resulting in an allocation that does not maximize utilization. Similarly, it is not perfectly fair, as once there are only a few tickets left, small groups still have a chance of being allocated but large groups do not. Thus, in the Group Lottery smaller groups are always weakly better than larger groups. This is formally stated in Lemma~\ref{lemma:gl-small-groups-advantage} located in Appendix~\ref{appendix:GL}.

\begin{theorem}[``GL is Good"]\label{thm:gl-is-good}
Fix $\kappa,\alpha \in (0,1)$. For every instance in $I(\kappa,\alpha)$, the group request equilibrium outcome of the Group Lottery is $(1-\kappa)$-efficient and $(1-2 \kappa)$-fair.
\end{theorem}

In Proposition~\ref{prop:gl-tightness} (located in Appendix~\ref{appendix:GL}), we construct instances where the fairness of the Group Lottery is arbitrarily close to the guarantee provided in Theorem~\ref{thm:gl-is-good}. These instances are fairly natural: groups all have size one or two, and the total number of tickets is odd. These conditions are met by the Hamilton Lottery, which we discuss in Section~\ref{sec:discussion}.

\subsubsection{Proof Sketch of Theorem~\ref{thm:gl-is-good}.}

The efficiency guarantee is based on the fact that for any order over groups, the number of tickets wasted can be at most $\max_G |G| - 1$. Therefore, the tight lower bound on efficiency of the group lottery is $1-\frac{\max_G |G|-1}{k} \ge 1 -\kappa$.\\

We now turn to the the fairness guarantee. We will show that for any pair of agent \(i,j\),
\begin{equation}\label{eq:gl-utility-ratio}
\frac{u_i(\GL(\a))}{u_j(\GL(\a))} \ge (1-2 \kappa).
\end{equation}
Because all groups are following the group request strategy, the set of valid groups is \(\G\). Fix an arbitrary agent \(i\). We construct a uniform random order over \(\G\) using Algorithm~\ref{alg:simultaneous}: first generate a uniform random order \(\rOrder^{-i}\) over \(\G\setminus G_i\), and then extend it to \(\G\) by uniformly inserting \(G_i\) in \(\rOrder^{-i}\). Moreover, if groups in \(\G\setminus G_i\) are processed according  \(\rOrder^{-i}\), then \(\tau(k - |G_i| + 1,\rOrder^{-i})\) represents the last step in which at least \(|G_i|\) tickets remains available. Therefore, if \(G_i\) is inserted in the first \(\tau(k - |G_i| + 1,\rOrder^{-i})\) positions it will get a payoff of \(1\). This is formalized in the next lemma.

\begin{lemma}\label{lemma:gl-utility-bounds}
For any instance in \(I(\kappa,\alpha)\) and any agent \(i\), if we let \(\a\) be the group request strategy under the Group Lottery and \(\rOrder^{-i}\) be a uniform order over \(\G\setminus G_i\), then
\begin{equation}
u_i(\GL(\a)) = \frac{\E[\tau(k - |G_i| + 1,\rOrder^{-i})]}{m} \le\frac{k}{n} \left(1 + \kappa\right),
\end{equation}
where \(\tau(k - |G_i| + 1,\rOrder^{-i})\) is as in \eqref{eq:def:tau} using the cardinality of each group as the size function.
\end{lemma}

Lemma~\ref{lemma:gl-small-groups-advantage} in Appendix~\ref{appendix:GL} states that if two groups are selecting the group request strategy under the Group Lottery, then the utility of the smaller group will be at least the utility of the larger group.  Therefore, we assume without loss of generality that \(|G_i| \ge |G_j|\). From Lemma~\ref{lemma:gl-utility-bounds}, it follows that
\begin{align}
\frac{u_i(\GL(\a))}{u_j(\GL(\a))}
& = \frac{\E[\tau(k-|G_i|+1,\rOrder^{-i})]}{\E[\tau(k-|G_j|+1,\rOrder^{-j})]} \ge \frac{\E[\tau(k-|G_i|+1,\rOrder^{-i})]}{\E[\tau(k-|G_j|+1,\rOrder^{-i})]}.
\end{align}
To complete the proof, we express the denominator on the right hand side as the sum of the numerator and the difference
\[\E[\tau(k-|G_j|+1,\rOrder^{-i})] - \E[\tau(k-|G_i|+1,\rOrder^{-i})], \]
which reflects the advantage of the small group \(G_j\). We bound this ratio by taking a lower bound on the numerator $\E[\tau(k-|G_i|+1,\rOrder^{-i})]$ and an upper bound on the difference $\E[\tau(k-|G_j|+1,\rOrder^{-i})] - \E[\tau(k-|G_i|+1,\rOrder^{-i})]$. Both bounds follow from the lemma below.

\begin{proposition}\label{lemma:mart-bounds}
Given a sequence of numbers \(\{a_1,\ldots,a_n\}\) such that \(a_t \ge 1\), define \(\mu = \sum_i a_i/n\) and \(\bar a = \max a_i\). Let \(\order\) be an order over \(\{1,\ldots,n\}\). For $k \in \{1, \ldots, \sum_i a_i\}$, we let \(\tau = \tau(k,\order)\) be as in \eqref{eq:def:tau} where the size of \(i\) is \(a_i\), that is, \(|\order_t| = a_{\order_t}\).
If \(\rOrder\) is a uniform random order of $\{1, \ldots, n\}$, then
\begin{equation}\label{eq:mart-bounds}
1 + \frac{k-\bar a}{\mu}\le \E[\tau(k,\rOrder)] \le \frac{k +\bar a -1}{\mu}.
\end{equation}
Furthermore, if \(k,k' \in \N\) are such that \(k + k' \le \sum_i a_i\) then
\begin{equation}\label{tau-diff}
\E[\tau(k',\rOrder)] +  \E[\tau(k,\rOrder)] \geq  \E[\tau(k' + k,\rOrder)].
\end{equation}
\end{proposition}

Equation \eqref{eq:mart-bounds} establishes that the expected time to reach $k$ is approximately $k$ divided by the average size $\mu$, while \eqref{tau-diff} establishes that hitting times are sub-additive. Both results are well known when the values \(a_{\rOrder_t}\) are sampled with replacement from \(\{a_1,\ldots,a_n\}\). Proposition \ref{lemma:mart-bounds} establishes the corresponding results when values are sampled {\em without} replacement. The proof of \eqref{eq:mart-bounds} employs a martingale presented in~\citet{serfling1974probability}, while the proof of \eqref{tau-diff} uses a clever coupling argument.\footnote{We thank Matt Weinberg for suggesting the appropriate coupling.} Although both statements are intuitive, we have not seen them proven elsewhere, and we view Proposition \ref{lemma:mart-bounds} as a statement whose importance extends beyond the setting in which we deploy it.

\subsubsection{A Fair Group Lottery.}

Theorem~\ref{thm:gl-is-good} establishes that the Group Lottery has strong performance guarantees. However, this mechanism is not perfectly fair, as small groups have an advantage over large groups, nor perfectly efficient, as the last few tickets might be wasted. It is natural to ask whether there exists a mechanism that overcomes these issues. Proposition~\ref{prop:badnews} shows that the best we can hope for is a mechanism that is $(1-\kappa)$-efficient and fair. We then describe a fairer version of the Group Lottery which attains these performance guarantees. We conclude with a discussion of advantages and disadvantages of this fair Group Lottery.

\begin{proposition}\label{prop:badnews}\hfill
\begin{enumerate}
    \item
    Fix $\kappa,\alpha \in (0,1)$. For every instance in $I(\kappa,\alpha)$, there exist a random allocation that is $(1-\kappa)$-efficient and fair.
    \item For any $\epsilon > 0$, there exists \(\kappa,\alpha \in (0,1)\) and an instance in \(I(\kappa, \alpha)\) such that no random allocation is
   $(1 - \kappa + \epsilon)$-efficient and $\epsilon$-fair.
\end{enumerate}
\end{proposition}

The first statement follows from a result in~\citet{nguyen2016assignment}, which implies that any utility vector such that (i) the sum of all agents' utilities is at most \(k - \max_{G \in \G}|G| + 1\), and (ii) members of the same group have identical utility, can be induced by a lottery over feasible allocations. To prove the second part of Proposition \ref{prop:badnews}, we construct an instance where a particular group must be awarded in order to avoid wasting a fraction \(\kappa\) of the tickets. Therefore, to improve beyond \((1-\kappa)\)-efficiency it is necessary to allocate that group more frequently. The complete proof of Proposition~\ref{prop:badnews} is located in Appendix~\ref{appendix:RA}.

Proposition~\ref{prop:badnews} establishes that the best guarantee we can hope for is a mechanism that is $(1-\kappa)$-efficient and fair. In fact, this can be achieved by first asking agents to identify their groups (as in the Group Lottery), and then allocating according to the random allocation referred to in the first part of Proposition~\ref{prop:badnews}. When using this mechanism, it is dominant for a group to truthfully report its members, as long as it can not influence the total number of tickets awarded.
In the following discussion, we refer to this mechanism as the \emph{Fair Group Lottery}.

To conclude this section, we discuss the trade-offs between the Fair Group Lottery and the Group Lottery. In light of its stronger performance guarantees, one might conclude that the Fair Group Lottery is superior. However, we think that there are several practical reasons to favor the standard Group Lottery. First, the computation of the Fair Group Lottery outcome is not trivial: \citet{nguyen2016assignment} give two procedures, one which they acknowledge might be ``impractical for large markets,'' and the other which returns only an approximately fair allocation. By contrast, the Group Lottery is simple to implement in code, and can even be run physically by writing applicants' names on ping-pong balls or slips of paper. In some settings, a physical implementation that allows applicants to witness the process may increase their level of trust in the system. Even when implemented digitally, the ability to explain the procedure to applicants may provide similar benefits.

A final benefit of the Group Lottery is that it provides natural robustness. Although we assume that the number of tickets is known in advance and that all successful applicants claim their tickets, either assumption may fail to hold in practice. When using the Group Lottery, if additional tickets become available after the initial allocation, they can be allocated by continuing down the list of groups. This intuitive policy preserves the fairness and efficiency guarantees from Theorem \ref{thm:gl-is-good}. By contrast, if tickets allocated by the Fair Group Lottery go unclaimed, there is no obvious ``next group" to offer them to, and any approach will likely violate the efficiency and fairness guarantees that this mechanism purports to provide.

For these reasons, we see the Group Lottery as a good practical solution in most cases: the \(2\kappa\) loss of fairness identified in Theorem  \ref{thm:gl-is-good} seems a modest price to pay for the benefits outlined above. There is a loose analogy to be drawn between the Fair Group Lottery and the Vickrey auction: although the Vickrey auction is purportedly optimal, practical considerations outside of the standard model prevent it from being widely deployed  \citep{ausubel2006lovely}. Similarly, a Fair Group Lottery is only likely to be used in settings satisfying several specific criteria: the institution running the lottery is both sophisticated and trusted, fairness is a primary concern, and applicants are unlikely to renege.

\subsection{Individual Lottery}

As noted in the introduction, asking for (and verifying) the identity of each participant may prove cumbersome. In this section we consider the widely-used {\em Individual Lottery}, in which the action set is \(A=\{1,\ldots,k\}\).\footnote{In practice, agents are often limited to asking for $\ell < k$ tickets. We refer to this mechanism as the {\em Individual Lottery with limit \(\ell\)}, and discuss it briefly at the end of the section. Appendix~\ref{appendix:IL} provides a complete analysis of this mechanism, and demonstrates that like the Individual Lottery without a limit, it can be arbitrarily unfair and inefficient.} Agents are placed in a uniformly random order and processed sequentially. Each agent is given a number of tickets equal to the minimum of their request and the number of remaining tickets.\footnote{As for the Group Lottery, one might imagine using a variant in which agents whose request exceeds the number of remaining tickets are skipped. The negative results in Theorem~\ref{thm:il-is-bad} would still hold when using this variant.}

More formally, given an action profile \(\a\in A^{\sN}\) and an order over agents \(\sigma \in \o_{\sN}\), we let \(\xil(\a,\sigma)\in \X\) be the feasible allocation  generated by the Individual Lottery:
\begin{equation}\label{eq:xil}
\xil_{\sigma_t}(\a,\sigma) = \min\left\{a_{\sigma_t}, \max\left\{k - \sum_{i \in \sigma_{[t-1]}} a_i,0\right\}\right\},
\end{equation}
for $t \in \{1,\ldots, n\}$.
For any \(x' \in \X\), the allocation function of the Individual Lottery is
\[\IL_{x'}(\a) = \frac{1}{n!}\sum_{\order \in \o_\sN}\ind{x' = \xil(\a,\order)}.\]

\subsubsection{Incentives.}

As in the Group Lottery, we refer to the strategy that correspond to truthful behavior as the group request strategy. In the Individual Lottery, this is the strategy in which each agent declares his or her group size.
\begin{definition}\label{def:IL-GR}
\em In the Individual Lottery, we say that group \(G\) follows the {\bf group request strategy} if \(a_i = |G|\) for all \(i \in G\).
\end{definition}

Our next result establishes that because agents' request do not affect the order in which they are processed, each agent should request at least his or her group size.
\begin{proposition}\label{prop:IL-incentives}
In the Individual Lottery, the set of actions \(\a_{G}\) is dominant for group \(G\) if and only if \(a_i \geq |G|\) for all \(i \in G\).
\end{proposition}

\subsubsection{Performance.}

Proposition~\ref{prop:IL-incentives} states that it is a dominant strategy to follow the group request strategy, but that there are other dominant strategies in which agents inflate their demand (select \(a_i > |G_i|\)).\footnote{Our model assumes that agents are indifferent between all allocations that allow all members of their group to receive a ticket. While we believe this to be a reasonable approximation, in practice, groups might follow the group request strategy if each ticket has a cost, or inflate their demand if tickets can be resold on a secondary market.} Our next result implies that the group request equilibrium Pareto dominates any other dominant strategy equilibrium.

\begin{proposition}\label{prop:IL-monotonicity}
Let \(i\) be any agent, fix any \(\a_{-i} \in \{1, 2, \ldots, k\}^{\sN \backslash \{i\}}\), and let \(a_i' > a_i \geq |G_i|\).  Then for every agent $j \in \sN$,
\[u_j(\IL(a_i,\a_{-i})) \ge u_j(\IL(a_i',\a_{-i})).\]
\end{proposition}

Even when agents request only the number of tickets needed by their group, the outcome will be inefficient if there are multiple winners from the same group. One might expect that this is unlikely if the supply/demand ratio $\alpha$ is small. However, Theorem~\ref{thm:il-is-bad} show that even in this case, the individual lottery can be arbitrarily unfair and inefficient.

\begin{theorem}[``IL is Bad'']\label{thm:il-is-bad}
For any \(\alpha, \kappa, \epsilon \in (0, 1)\), there exists an instance in \(I(\kappa,\alpha)\) such that any dominant strategy equilibrium outcome of the Individual Lottery is not \(\epsilon-\)efficient nor \(\epsilon-\)fair.
\end{theorem}

\subsubsection{Proof Sketch of Theorem~\ref{thm:il-is-bad}.}

We will construct an instance in \(I(\kappa, \alpha)\), where the outcome of the Individual Lottery is arbitrarily unfair and arbitrarily inefficient. In this instance, there are \(n\) agents and \(k = \alpha n\) tickets. Furthemore, agents are divided into one large group of size \(n^{3/4}\) and \(n - n^{3/4}\) groups of size one. If \(n\) is large enough, then this instance is in \(I(\kappa, \alpha)\) and the following two things happen simultaneously:

\begin{enumerate}
    \item The size of the large group \(n^{3/4}\) is small relative to the number of tickets \(k = \alpha n\).
    \item The fraction of tickets allocated to small groups is insignificant.
\end{enumerate}
Hence, the resulting allocation is unfair as the large group has an advantage over small groups, and inefficient as a vanishing fraction of the agents get most of the tickets.

Formally, let agents \(i,j\) be such that \(|G_i| = 1\) and \(|G_j| = n^{3/4}\). We will start by proving the efficiency guarantee. By Proposition~\ref{prop:IL-monotonicity} it follows that the group request is the most efficient dominant action profile. Therefore, we assume without loss of generality that this action profile is being selected. The utilization of this system is
\begin{equation}\label{eq:il-bad-instance-utilization}
\frac{n^{3/4}  u_j(\IL(\a))}{k} + \frac{(n -  n^{3/4}) u_i(\IL(\a))}{k}.
\end{equation}
We now argue that both terms in~\eqref{eq:il-bad-instance-utilization} can be made arbitrarily small by making \(n\) sufficiently large. We begin by studying the first term in~\eqref{eq:il-bad-instance-utilization}. Using the fact that utilities are upper bounded by \(1\), it follows that
\[\frac{n^{3/4}  u_j(\IL(\a))}{k} \le \frac{n^{3/4}}{k} = \frac{1}{\alpha n^{1/4}}.\]
Hence, to ensure that \((1)\) holds it suffices to have \(n\) growing to infinity. We now analyze the second term in~\eqref{eq:il-bad-instance-utilization}. Because the group request action profile is being selected, this term is equal to the fraction of tickets allocated to small groups.

Moreover, we show the following upper bound on utility of agent \(i\):
\begin{equation}\label{eq:il-large-group-ub-body}
u_i(\IL(\a)) \le \frac{k}{(n^{3/4})^2} = \frac{\alpha}{n^{1/2}}.
\end{equation}
The intuition behind this bound is as follows. If we restrict our attention only to agents in \(G_i\) and \(G_j\), then we know that \(i\) will get a payoff \(0\) unless it is processed after at most \(k/n^{3/4} - 1\) members of \(G_j\). Because the order over agents is uniformly distributed, this event occurs with probability
\[\frac{k/n^{3/4}}{n^{3/4} + 1} \le \frac{k}{(n^{3/4})^2}.\]
From the first inequality in~\eqref{eq:il-large-group-ub-body}, it follows that
\[\frac{(n -  n^{3/4}) u_i(\IL(\a))}{k} \le \frac{n}{n^{3/2}} \le \frac{1}{n^{1/2}}.\]
Notice that the right hand side goes to \(0\) as \(n\) grows, so \((2)\) holds.

We now turn to the fairness guarantee. To this end, we use a trivial upper bound on the utility of agent \(j\), based on the fact that the first agent to be processed always obtains a payoff of \(1\). Thus,
\begin{equation}%\label{eq:il-large-group-ub-body}
u_j(\IL(\a)) \ge \frac{n^{3/4}}{n} = n^{-1/4}.
\end{equation}
Note that this lower bound is attained when all agents in small groups request \(k\) tickets. Combining the bound above and the second inequality in~\eqref{eq:il-large-group-ub-body}, we obtain
\begin{equation}%\label{eq:il-large-group-ub-body}
\frac{u_i(\IL(\a))}{u_j(\IL(\a))} \le \frac{\alpha n^{1/4}}{n^{1/2}}.
\end{equation}

We conclude noting that the right side goes to \(0\) as \(n\) grows. The full proof of Theorem~\ref{thm:il-is-bad} is located in Appendix~\ref{appendix:IL}.

\subsubsection{Limiting the Number of Tickets Requested.}

In many applications, a variant of the Individual Lottery is used where a limit is imposed on the number of tickets an agent can request. For example, in the Hamilton Lottery agents can request at most \(2\) tickets, and in the Big Sur Marathon groups can have at most 15 individuals. This motivates us to study the Individual Lottery with limit \(\ell\). Formally, the only difference between this and the standard Individual Lottery is the action set, which is \(A = \{1,\ldots,\ell\}^n\), with the limit \(\ell\) chosen by the designer.

The choice of limit must balance several risks. Imposing a limit of $\ell$ reduces the risk from inflated demand, but harms groups with more than $\ell$ members. The latter effect reduces fairness and may also reduce efficiency if there are many large groups. In fact, we show in Proposition~\ref{prop:il-ell-bad} that the Individual Lottery with limit \(\ell\) is still arbitrarily unfair and arbitrarily inefficient in the worst case.
\begin{proposition}\label{prop:il-ell-bad}
For any \(\alpha, \kappa, \epsilon \in (0, 1)\) and \(\ell \in \N\), there exists an instance in \(I(\kappa,\alpha)\) such that, regardless the action profile selected, the outcome of the Individual Lottery with limit \(\ell\) is not \(\epsilon-\)efficient nor \(\epsilon-\)fair.
\end{proposition}

The proof of Proposition~\ref{prop:il-ell-bad} is in Appendix~\ref{appendix:IL}. In the example considered in the proof, problems stem from the fact that most groups have more than $\ell$ members. However, even if group sizes are upper bounded by \(\ell\), the Individual Lottery with limit \(\ell\) still performs poorly in the worst case. In particular, Propositions~\ref{prop:il(ell)-eff} and~\ref{prop:il(ell)-fairness} show that, if no group have more than \(\ell\) members, every dominant strategy equilibrium outcome of the Individual Lottery with limit \(\ell\) is \(1/\ell-\)efficient and \(1/\ell\)-fair.  Moreover, these guarantees are tight in the worst case. We give a complete analysis of this variant of the Individual Lottery  in Appendix~\ref{appendix:IL}.

\subsection{\NameProposedMechanism}

The example presented in Theorem~\ref{thm:il-is-bad} is an extreme case that we shouldn't see too often in practice. However, it illustrates the major issues of the Individual Lottery. In this section, we show that minor modifications to the Individual Lottery can yield strong performance guarantees even in these extreme cases.

We study the {\NameProposedMechanism} ($IW$), whose only departure from the Individual Lottery is the order in which agents are placed. Instead of using a uniform random order, the {\NameProposedMechanism} uses a random order biased against agents with large requests. Theorem~\ref{thm:spl-performance} shows that the {\NameProposedMechanism} is approximately fair and approximately efficient, and similar to the Group Lottery when there are many more agents than tickets.

Formally, each agent selects an action in $\{1,....,k\}$. For each \(\order \in \o_{\sN}\), we let random order over agents $\rOrder$ be such that

\begin{equation}\label{eq:size_proportional_lottery}
\bP(\rOrder=\order \vert \a) = \prod_{t=1}^{n}\frac{1/a_{\order_t}}{\sum \limits_{i \in N \backslash \order_{[t-1]} }1/a_{\order_i}}.
\end{equation}

There are several ways to generate \(\rOrder\). This order can be thought of as the result of sequentially sampling agents without replacement, with probability inversely proportional to the number of tickets that they request. One property that motivates the study of the {\NameProposedMechanism} is that when agents declare their group size, every group that has not been drawn is equally likely to be draw next.

Let $\rOrder \in \o_{\sN}$ be distributed according to \eqref{eq:size_proportional_lottery}. For any \(x' \in \X\), the allocation function of the {\NameProposedMechanism} is
\[\SPL_{x'}(\a) = \sum_{\order \in \o_\sN}\ind{x' = \xil(\a,\order)}\bP(\rOrder = \order),\]
with $\xil$ defined as in~\eqref{eq:xil}.

We define group request strategy as in Definition~\ref{def:IL-GR}: each agent $i$ requests $|G_i|$ tickets.

\subsubsection{Incentives.}

In this section, we will see that under the {\NameProposedMechanism}, there are instances where no strategy is dominant for every group. However, we will argue that if demand significantly exceeds supply, then it is reasonable to assume that groups will select the group request strategy.

We start by showing in Proposition~\ref{prop:SPL-incentives}, that for groups of size three or less the group request is the only dominant strategy.

\begin{proposition}\label{prop:SPL-incentives}
In the {\NameProposedMechanism}, if $G\in\G$ is such that $|G| \leq 3$, then the group request is the only dominant strategy for $G$.
\end{proposition}

The following example shows that for groups of more than three agents, deviating from the group request is potentially profitable.

\begin{example}\label{ex:SPL-incentives}
Consider an instance with \(n\) agents and \(n-1\) tickets. We divide the agents into one group of size $4$ and $n-4$ groups of a single agent. If \(n \ge 17\), then the optimal strategy for the large group will depend on the action profile selected by the small groups. In particular, if the small groups are following the group request strategy, then members of the larger group benefit from each requesting $2$ tickets instead of $4$. The analysis of this example is located in Appendix~\ref{app:spl}.
\end{example}

In the example above, when \(n \le 16\), it is actually optimal for the large group to play the group request. Thus, this deviation is only profitable when \(n \ge 17\), and the group success probability is bigger than \(92\%\). In general, when agents request fewer tickets than their group size, their chance of being selected increases, but now multiple agents from the group must be drawn in order to achieve success. This should be profitable only if the chance of each agent being drawn is high.

We formalize this intuition in Conjecture~\ref{conj:spl-incentives}. Roughly speaking, we conjecture that in scenarios where the success probability of a group is below $1 - 1/e \approx 63\%$, the group request strategy maximizes its conditional expected utility. Proposition~\ref{prop:spl-gr-optimality-restricted} lends additional support to the conjecture. This proposition establishes that our conjecture holds when restricted to a broad set of strategies. In order to present our conjecture, we need first to introduce some definitions.

In what follows, we fix an arbitrary group \(G\). Given any action profile \(\a\), we generate an order over agents $\rOrder$ using the following algorithm:

\begin{algorithm}[H]
\caption{\label{alg:exponentials}}
\begin{enumerate}
\item Draw $\{X_i\}_{i\in \sN}$ as i.i.d. exponentials, with $\bP(X_i > t) = e^{-t}$ for $t \geq 0$.
\item Place agents in increasing order of $a_i X_i$: that is, output $\Sigma$ such that
\[a_{\Sigma_1}X_{\Sigma_1} < \cdots < a_{\Sigma_n}X_{\Sigma_n}.\]
\end{enumerate}
\end{algorithm}

From Proposition~\ref{prop:spl-random-order} in Appendix~\ref{app:spl}, it follows that \(\rOrder\) is distributed according to~\eqref{eq:size_proportional_lottery} conditional on $\a$. We will refer to \(a_i X_i\) as the {\em score} obtained by agent \(i\). Note that a lower score is better as it increases the chances of getting awarded.

The usual way to study the incentives of group \(G\), is to find a strategy that maximizes its utility given the actions of other agents. Here, we will assume that \(G\) has an additional information: the scores of other agents. Thus, we study the problem faced by \(G\) of maximizing its success probability given actions and scores of everyone else. This problem seems to be high-dimensional and very complex, however, we will show that all the information relevant for \(G\) can be captured by a sufficient statistic \(T\). Define,

\begin{equation}\label{eq:T-def}
T = \inf\left\{t \in \R: \sum_{j \not\in G} a_j \ind{a_j X_j < t} > k - |G|\right\}.
\end{equation}

We show in Lemma~\ref{lemma:threshold} located in Appendix~\ref{app:spl}, that \(G\) gets a utility of \(1\) if and only if the sum of the requests of its members whose score is lower than \(T\) is at least \(|G|\). Therefore, we can formulate the problem faced by \(G\) as follows:

\begin{equation}\label{eq:spl-GT-problem}
\begin{array}{rlll}
\max &\bP(\sum_{i\in G} a_i\ind{a_i X_i < T} \ge |G|)\\
 \subjectto &a_i \in \{1,\ldots,k\}\quad \forall i \in G.
\end{array}
\end{equation}

Notice that under the group request strategy, the objective value in~\eqref{eq:spl-GT-problem} evaluates to

\begin{equation}\label{eq:spl-GT-problem-gr}
1 - e^{-T}.
\end{equation}

This follows because \(G\) will get a payoff of \(1\) if and only if at least one of its members has a score lower than \(T\), that is, \(\min_{i\in G}\{a_i X_i\} < T\). Moreover, using that \(\sum_{i\in G} 1/a_i = 1\), and by the well-known properties of the minimum of exponential random variables, we have that $\min_{i\in G}\{a_i X_i\} \sim Exp(1)$.

Definitions out of the way, we can present our conjecture and the proposition supporting it.

\begin{conjecture}\label{conj:spl-incentives}If \(T \le 1\), then no other strategy yields a higher objective value in~\eqref{eq:spl-GT-problem} than the group request.
\end{conjecture}

From equation~\eqref{eq:spl-GT-problem-gr}, we see that the utility yield by the group request is an increasing function of \(T\). Therefore, we can think of \(T\) as an indicator of how competitive the market is. Thus, the interpretations of Conjecture~\ref{conj:spl-incentives} is that if the market is moderately competitive ($G$ has success probability below $1 - 1/e \approx 63\%$), then the group request is optimal. While we haven't proved the conjecture, we do have a proof that it holds for a broad subset of strategies. For \(r \in \{0,\ldots, |G|-1\}\), we define \(\B_r \subseteq A_G\) to be the set of strategies for which the sum of any \(r\) requests is less than \(|G|\), while the sum of any \(r+1\) requests is greater than or equal to \(|G|\).  Let

\begin{equation}
\B = \bigcup_{r=0}^{|G|-1} \B_r.
\end{equation}

\begin{proposition}\label{prop:spl-gr-optimality-restricted}
If \(T \le 1\), then no other strategy in \(\B\) yields a higher objective value in~\eqref{eq:spl-GT-problem} than the group request.
\end{proposition}

Note that \(\B\) is rich enough such that for any group of size greater than \(3\), it contains a strategy that is better than the group request for \(T\) large enough.

\proof[Proof sketch of Proposition~\ref{prop:spl-gr-optimality-restricted}]
In this proof, we will say that an agent is awarded if and only if it has a score lower than \(T\).

From~\eqref{eq:spl-GT-problem-gr}, it suffices to show that under any strategy in \(\B_r\), the objective value in~\eqref{eq:spl-GT-problem} is at most \(1-e^{-T}\).
We start by studying a relaxation of the problem defined in~\eqref{eq:spl-GT-problem}.  In this relaxation, the number of times an agent \(i\) is awarded follows a Poisson distribution with rate \(T/a_i\), and the total number of times \(G\) is awarded follows a Poisson distribution with rate \(\sum_{i\in G}T/a_i\). Note that if the set of feasible strategies is \(\B_r\), then by Lemma~\ref{lemma:threshold} in Appendix~\ref{app:spl} it follows that \(G\) needs to be awarded at least \(r+1\) times. Finally, using a Poisson tail bound, we show that this event happens with probability at most \(1 -e^{-T}\). This bound can only be applied if the expected number of times \(G\) is awarded is at most \(r+1\). This follows because \(T\le 1\) and Lemma~\ref{lemma:spl-reciprocals-ub} in Appendix~\ref{app:spl}, which establishes that for any \(\a_G' \in \B_r,~\sum_{i\in G}1/a'_i \le r+1\).
\endproof

\subsubsection{Performance.}

We now study the performance of the {\NameProposedMechanism}, under the assumption that groups are selecting the group request strategy. We think that this assumption is reasonable for two reasons: (i) for groups of size at most three, the group request is the only dominant strategy, and (ii) for larger groups, we conjecture that in scenarios where its success probability is moderate (at most \(63\%\)), the group request strategy is optimal. The main result of this section is Theorem~\ref{thm:spl-performance}, which establishes that the {\NameProposedMechanism} is approximately efficient and fair.

To state these guarantees, we define for any \(x>0\),

\begin{equation}
    g(x) = \frac{1-e^{-x}}{x}.
\end{equation}

\begin{theorem}\label{thm:spl-performance}
 Fix $\kappa,\alpha \in (0,1)$. For every instance in $I(\kappa,\alpha)$, the group request outcome of the {\NameProposedMechanism} is $(1 - \kappa)g(\alpha)$-efficient and \((1-2\kappa)g(\alpha)\)-fair.
\end{theorem}

These guarantees resemble the ones offered for the Group lottery. Recall that Theorem~\ref{thm:gl-is-good} establishes that the Group Lottery is \((1-\kappa)\)-efficient and \((1-2\kappa)\)-fair. It is not perfectly efficient, as the last tickets might be wasted. Similarly, it is not perfectly fair, as once there are only a few tickets left, small groups still have a chance of being allocated but large groups do not. These issues persist under the {\NameProposedMechanism}. In addition, the {\NameProposedMechanism} has the additional concern that multiple member of a group may be selected. This explains the multiplicative factor of  \(g(\alpha)\) in the theorem statement. Because \(g(\alpha) \ge 1 - \alpha/2\), when \(\alpha\) is close to \(0\) the guarantees for the Group Lottery and the {\NameProposedMechanism} coincide. Although it is intuitive that a small supply-demand ratio implies a small chance of having groups with multiple winners, the previous section shows that this may not be the case under the standard Individual Lottery.

\subsubsection{Proof Sketch of Theorem~\ref{thm:spl-performance}.}

In order to prove the efficiency and fairness guarantees, we first introduce a new mechanism: the {\em Group Lottery with Replacement (GR).} This is a variant of the Group Lottery in which valid groups can be processed more than once. Formally, the set of actions, the set of valid groups \(V\), the group request strategy and the allocation rule \(\xgl\) are defined exactly as in the Group Lottery. However, the allocation function \(\GLR\) is different, in particular, this mechanism process valid groups according to a sequence of \(k\) elements \(\sq \in \s_V\) , where \(\sq_t\) is independently and uniformly sampled with replacement from \(V\).
Hence, for any \(x' \in \X\), the allocation function of the Group Lottery is
\[\GLR_{x'}(\a) = \sum_{\order \in \o_V}\ind{x' = \xgl(\a,\order)}\bP(\sq = \order),\]
with $\xgl$ defined as in~\eqref{eq:xgl}.
Having defined this new mechanism, we now present a lemma that will be key in proving both guarantees. This lemma establishes a dominance relation between the {\NameProposedMechanism}, the Group Lottery and the Group Lottery with Replacement, when the group request action profile is being selected. As we will see, every agent prefers the Group Lottery to the {\NameProposedMechanism}, and  the {\NameProposedMechanism} to the Group Lottery with Replacement.
\begin{lemma}\label{prop:mech-dominance}
For any instance and any agent \(i\in \sN\), if \(\a\) denote the corresponding group request action profile for each mechanism below, then
\begin{equation}\label{eq:mech-dominance}
u_i(\GLR(\a)) \le u_i(\SPL(\a)) \le u_i(\GL(\a)).
\end{equation}
\end{lemma}
The key idea to prove Lemma~\ref{prop:mech-dominance} is that the order or sequence used in each of these mechanism can be generated based on a random sequence of agents \(\sq'\). Roughly speaking, each order or sequence is generated from \(\sq'\) as follows:
\begin{itemize}
    \item Group Lottery with replacement: replace every agent %\(\sq'_t\)
    by its group. %\(G_{\sq'_t}\).
    \item \NameProposedMechanism: remove every agent that has already appeared in a previous position.
    \item Group Lottery: replace every agent by its group, and then remove every group that has already appeared in a previous position.
\end{itemize}
Note that because in each mechanism the group request strategy is being selected, whenever a group or agent is being processed, it is given a number of tickets equal to the minimum of its group size and the number of remaining tickets.

This implies that, under the Group Lottery with Replacement, a group could be given more tickets than needed because one of its members appeared more than once in the first positions of \(\sq'\). This situation is avoided in the {\NameProposedMechanism}, hence, making all agents weakly better. Similarly, under the {\NameProposedMechanism}, a group could be given more tickets than needed because its members appeared more than once in the first positions of \(\sq'\). This situation is avoided in the Group Lottery, hence, making all agents weakly better. The full proof is located in Appendix~\ref{app:mech-dominance}.

We now turn to the efficiency guarantee. From Lemma~\ref{prop:mech-dominance}, it follows that for any instance the utilization under the {\NameProposedMechanism} is at least the utilization under the Group Request with Replacement.  Therefore, it suffices to show that for any instance in \(I(\kappa,\alpha)\), the Group Lottery with Replacement is \((1 - \kappa)g(\alpha)\)-efficient. To this end, we present in Lemma~\ref{lemma:glr-utility-lb-body} a lower bound on the utility of any agent under the Group Lottery with Replacement.
\begin{lemma}\label{lemma:glr-utility-lb-body}
For any instance in \(I(\kappa,\alpha)\) and any agent \(i\), if we let \(\a\) be the group request under the Group Lottery with Replacement, then \vspace{-.1 in}
\begin{equation}\label{eq:glr-utility-lb}
u_i(\GLR(\a)) \ge \frac{k}{n}(1 - \kappa) g(\alpha).
\end{equation}
\end{lemma}
The proof of Lemma~\ref{lemma:glr-utility-lb-body} is in Appendix~\ref{app:glr}. This lemma immediately give us the desired lower bound on the utilization of the Group Lottery with replacement.

We now show the fairness guarantee. From Lemma~\ref{prop:mech-dominance}, we have that for any instance and any pair of agents \(i,j\),
\begin{equation}\label{eq:GL-utility-ratios}
\frac{u_i(\SPL(\a))}{u_j(\SPL(\a))} \ge \frac{u_i(\GLR(\a))}{u_j(\GL(\a))}.
\end{equation}
Hence, it suffices to show that the ratio on the right hand side is at least \((1-2\kappa)g(\alpha)\). In Lemma~\ref{lemma:gl-utility-bounds} we proved an upper bound on the utility of an agent under the Group Lottery. Meanwhile, in Lemma~\ref{lemma:glr-utility-lb-body} we established a lower bound on the utility of an agent under the Group Lottery with Replacement.  Combining equation~\eqref{eq:GL-utility-ratios}, Lemma~\ref{lemma:gl-utility-bounds} and Lemma~\ref{lemma:glr-utility-lb-body} yields our fairness factor of \((1-2\kappa)g(\alpha)\).

\section{Discussion}\label{sec:discussion}

We consider a setting where groups of people wish to share an experience that is being allocated by lottery. We study the efficiency and fairness of simple mechanisms in two scenarios: one where agents identify the members of their group, and one where they simply request a number of tickets. In the former case, the Group Lottery is \((1-\kappa)\)-efficient and \((1-2\kappa)\)-fair. However, its natural and widespread counterpart, the Individual Lottery, suffers from deficiencies that can cause it to be arbitrarily inefficient and unfair. As an alternative, we propose the Weighted Individual Lottery. This mechanism uses the same user interface as the Individual Lottery, and Theorem~\ref{thm:spl-performance} establishes that it is \((1-\kappa)g(\alpha)\)-efficient and \((1-2\kappa)g(\alpha)\)-fair.

Although our bounds are based on worst-case scenarios, they can be combined with publicly available data to provide meaningful guarantees. In 2016 the Hamilton Lottery received approximately $n = 10,000$ applications daily for $k = 21$ tickets, with a max group size of $s = 2$.\footnote{Source: \url{https://www.bustle.com/articles/165707-the-odds-of-winning-the-hamilton-lottery-are-too-depressing-for-words}.} Hence, in this case \(\kappa \le .05\) and \(g(\alpha) \ge .99\). Therefore, by Theorem~\ref{thm:gl-is-good}, the Group Lottery outcome is at least $95\%$ efficient and $90\%$ fair. Furthermore, Theorem~\ref{thm:spl-performance} gives nearly identical guarantees for the {\NameProposedMechanism}. Meanwhile, the 2020 Big Sur Marathon Groups and Couples lottery received $1296$ applications for $k = 702$ tickets, with a maximum group size of $15$.\footnote{Source: \url{https://www.bigsurmarathon.org/random-drawing-results-for-the-2020-big-sur-marathon/}.} This yields \(\kappa = 14/702 \approx .02\), so Theorem \ref{thm:gl-is-good} implies $98\%$ efficiency and $96\%$ fairness for the Group Lottery. Determining $\alpha$ is tricker, as we do not observe the total number of agents $n$. Using the very conservative lower bound \(n \ge 1296\) (based on the assumption that every member of every group submitted a separate application) yields \(g(\alpha) \ge .77\). Thus, Theorem \ref{thm:spl-performance} implies $76\%$ efficiency and $74\%$ fairness for the {\NameProposedMechanism}. Its true performance would likely be much better, but accurate estimates would rely on understanding how many groups are currently submitting multiple applications.

Our analysis makes the strong assumption of dichotomous preferences. In practice, the world is more complicated: groups may benefit from extra tickets that can be sold or given to friends, and groups that don't receive enough tickets for everyone may choose to split up and have a subset attend the event. Despite these considerations, we believe that dichotomous preferences capture the first-order considerations in several markets while maintaining tractability. The mechanisms we have proposed, while imperfect, are practical and offer improvements over the Individual Lottery (which is often the status quo). Furthermore, we conjecture that the efficiency and fairness of the Group Lottery would continue to hold in a model where the utility of members of group $G$ is a more general function of the number of tickets received by $G$, so long as this function is convex and increasing on $\{1, 2, \ldots, |G|\}$, and non-increasing thereafter.

One exciting direction for future work is to adapt these mechanisms to settings with heterogeneous goods. For example, while the daily lottery for Half Dome allocates homogeneous permits using the Individual Lottery, the pre-season lottery allocates 225 permits for each day. Before the hiking season, each applicant enters a number of permits requested (up to a maximum of six), as well as a ranked list of dates that would be feasible. Applicants are then placed in a uniformly random order, and sequentially allocated their most preferred feasible date. This is the natural extension of the Individual Lottery to a setting with with heterogeneous goods, and has many of the same limitations discussed in this paper. It would be interesting to study the performance of (generalizations of) the Group Lottery and {\NameProposedMechanism} in this setting.

% In the interest of anonymization, please do not include acknowledgements in your submission.
%
%\begin{acks}
%
%	The authors would like to thank Dr. Maura Turolla of Telecom
%	Italia for providing specifications about the application scenario.
%
%	The work is supported by the \grantsponsor{GS501100001809}{National
%		Natural Science Foundation of
%		China}{http://dx.doi.org/10.13039/501100001809} under Grant
%	No.:~\grantnum{GS501100001809}{61273304\_a}
%	and~\grantnum[http://www.nnsf.cn/youngscientsts]{GS501100001809}{Young
%		Scientsts' Support Program}.
%
%
%\end{acks}

% Bibliography
\bibliographystyle{ACM-Reference-Format}
\bibliography{lotteries}


\begin{thebibliography}{13}

%%% the Web address.

\ifx \showCODEN    \undefined \def \showCODEN     #1{\unskip}     \fi
\ifx \showDOI      \undefined \def \showDOI       #1{#1}\fi
\ifx \showISBNx    \undefined \def \showISBNx     #1{\unskip}     \fi
\ifx \showISBNxiii \undefined \def \showISBNxiii  #1{\unskip}     \fi
\ifx \showISSN     \undefined \def \showISSN      #1{\unskip}     \fi
\ifx \showLCCN     \undefined \def \showLCCN      #1{\unskip}     \fi
\ifx \shownote     \undefined \def \shownote      #1{#1}          \fi
\ifx \showarticletitle \undefined \def \showarticletitle #1{#1}   \fi
\ifx \showURL      \undefined \def \showURL       {\relax}        \fi
% invisible to TeX
\providecommand\bibfield[2]{#2}
\providecommand\bibinfo[2]{#2}
\providecommand\natexlab[1]{#1}
\providecommand\showeprint[2][]{arXiv:#2}

\bibitem[\protect\citeauthoryear{Ausubel, Milgrom, et~al\mbox{.}}{Ausubel
  et~al\mbox{.}}{2006}]%
        {ausubel2006lovely}
\bibfield{author}{\bibinfo{person}{Lawrence~M Ausubel}, \bibinfo{person}{Paul
  Milgrom}, {et~al\mbox{.}}} \bibinfo{year}{2006}\natexlab{}.
\newblock \showarticletitle{The lovely but lonely Vickrey auction}.
\newblock \bibinfo{journal}{\emph{Combinatorial auctions}}
  \bibinfo{volume}{17} (\bibinfo{year}{2006}), \bibinfo{pages}{22--26}.
\newblock


\bibitem[\protect\citeauthoryear{Benassy}{Benassy}{1982}]%
        {benassy_1982}
\bibfield{author}{\bibinfo{person}{Jean~Pascal Benassy}.}
  \bibinfo{year}{1982}\natexlab{}.
\newblock \bibinfo{booktitle}{\emph{The economics of market disequilibrium}}.
\newblock \bibinfo{publisher}{New York: Academic Press}.
\newblock


\bibitem[\protect\citeauthoryear{Bogomolnaia and Moulin}{Bogomolnaia and
  Moulin}{2004}]%
        {bogomolnaia2004random}
\bibfield{author}{\bibinfo{person}{Anna Bogomolnaia} {and}
  \bibinfo{person}{Herv{\'e} Moulin}.} \bibinfo{year}{2004}\natexlab{}.
\newblock \showarticletitle{Random matching under dichotomous preferences}.
\newblock \bibinfo{journal}{\emph{Econometrica}} \bibinfo{volume}{72},
  \bibinfo{number}{1} (\bibinfo{year}{2004}), \bibinfo{pages}{257--279}.
\newblock


\bibitem[\protect\citeauthoryear{Bogomolnaia, Moulin, Stong,
  et~al\mbox{.}}{Bogomolnaia et~al\mbox{.}}{2005}]%
        {bogomolnaia2005collective}
\bibfield{author}{\bibinfo{person}{Anna Bogomolnaia},
  \bibinfo{person}{Herv{\'e} Moulin}, \bibinfo{person}{Richard Stong},
  {et~al\mbox{.}}} \bibinfo{year}{2005}\natexlab{}.
\newblock \showarticletitle{Collective choice under dichotomous preferences}.
\newblock \bibinfo{journal}{\emph{Journal of Economic Theory}}
  \bibinfo{volume}{122}, \bibinfo{number}{2} (\bibinfo{year}{2005}),
  \bibinfo{pages}{165--184}.
\newblock


\bibitem[\protect\citeauthoryear{Cachon and Lariviere}{Cachon and
  Lariviere}{1999}]%
        {cachon1999capacity}
\bibfield{author}{\bibinfo{person}{G{\'e}rard~P Cachon} {and}
  \bibinfo{person}{Martin~A Lariviere}.} \bibinfo{year}{1999}\natexlab{}.
\newblock \showarticletitle{Capacity choice and allocation: Strategic behavior
  and supply chain performance}.
\newblock \bibinfo{journal}{\emph{Management science}} \bibinfo{volume}{45},
  \bibinfo{number}{8} (\bibinfo{year}{1999}), \bibinfo{pages}{1091--1108}.
\newblock


\bibitem[\protect\citeauthoryear{Ching}{Ching}{1992}]%
        {ching_1992}
\bibfield{author}{\bibinfo{person}{Stephen Ching}.}
  \bibinfo{year}{1992}\natexlab{}.
\newblock \showarticletitle{A simple characterization of the uniform rule}.
\newblock \bibinfo{journal}{\emph{Economics letters}} \bibinfo{volume}{40},
  \bibinfo{number}{1} (\bibinfo{year}{1992}), \bibinfo{pages}{57--60}.
\newblock


\bibitem[\protect\citeauthoryear{Ehlers and Klaus}{Ehlers and Klaus}{2003}]%
        {ehlers2003probabilistic}
\bibfield{author}{\bibinfo{person}{Lars Ehlers} {and} \bibinfo{person}{Bettina
  Klaus}.} \bibinfo{year}{2003}\natexlab{}.
\newblock \showarticletitle{Probabilistic assignments of identical indivisible
  objects and uniform probabilistic rules}.
\newblock \bibinfo{journal}{\emph{Review of Economic Design}}
  \bibinfo{volume}{8}, \bibinfo{number}{3} (\bibinfo{year}{2003}),
  \bibinfo{pages}{249--268}.
\newblock


\bibitem[\protect\citeauthoryear{Johnson, Kemp, and Kotz}{Johnson
  et~al\mbox{.}}{2005}]%
        {johnson2005univariate}
\bibfield{author}{\bibinfo{person}{Norman~L Johnson},
  \bibinfo{person}{Adrienne~W Kemp}, {and} \bibinfo{person}{Samuel Kotz}.}
  \bibinfo{year}{2005}\natexlab{}.
\newblock \bibinfo{booktitle}{\emph{Univariate discrete distributions}}.
  Vol.~\bibinfo{volume}{444}.
\newblock \bibinfo{publisher}{John Wiley \& Sons}.
\newblock


\bibitem[\protect\citeauthoryear{Nguyen, Peivandi, and Vohra}{Nguyen
  et~al\mbox{.}}{2016}]%
        {nguyen2016assignment}
\bibfield{author}{\bibinfo{person}{Th{\`a}nh Nguyen}, \bibinfo{person}{Ahmad
  Peivandi}, {and} \bibinfo{person}{Rakesh Vohra}.}
  \bibinfo{year}{2016}\natexlab{}.
\newblock \showarticletitle{Assignment problems with complementarities}.
\newblock \bibinfo{journal}{\emph{Journal of Economic Theory}}
  \bibinfo{volume}{165} (\bibinfo{year}{2016}), \bibinfo{pages}{209--241}.
\newblock


\bibitem[\protect\citeauthoryear{Patel, Khan, and Louis}{Patel
  et~al\mbox{.}}{2020}]%
        {patel2020group}
\bibfield{author}{\bibinfo{person}{Deval Patel}, \bibinfo{person}{Arindam
  Khan}, {and} \bibinfo{person}{Anand Louis}.} \bibinfo{year}{2020}\natexlab{}.
\newblock \showarticletitle{Group Fairness for Knapsack Problems}.
\newblock \bibinfo{journal}{\emph{arXiv preprint arXiv:2006.07832}}
  (\bibinfo{year}{2020}).
\newblock


\bibitem[\protect\citeauthoryear{Roth, S{\"o}nmez, and {\"U}nver}{Roth
  et~al\mbox{.}}{2005}]%
        {roth2005pairwise}
\bibfield{author}{\bibinfo{person}{Alvin~E Roth}, \bibinfo{person}{Tayfun
  S{\"o}nmez}, {and} \bibinfo{person}{M~Utku {\"U}nver}.}
  \bibinfo{year}{2005}\natexlab{}.
\newblock \showarticletitle{Pairwise kidney exchange}.
\newblock \bibinfo{journal}{\emph{Journal of Economic theory}}
  \bibinfo{volume}{125}, \bibinfo{number}{2} (\bibinfo{year}{2005}),
  \bibinfo{pages}{151--188}.
\newblock


\bibitem[\protect\citeauthoryear{Serfling}{Serfling}{1974}]%
        {serfling1974probability}
\bibfield{author}{\bibinfo{person}{Robert~J Serfling}.}
  \bibinfo{year}{1974}\natexlab{}.
\newblock \showarticletitle{Probability inequalities for the sum in sampling
  without replacement}.
\newblock \bibinfo{journal}{\emph{The Annals of Statistics}}
  (\bibinfo{year}{1974}), \bibinfo{pages}{39--48}.
\newblock


\bibitem[\protect\citeauthoryear{Sprumont}{Sprumont}{1991}]%
        {sprumont_1991}
\bibfield{author}{\bibinfo{person}{Yves Sprumont}.}
  \bibinfo{year}{1991}\natexlab{}.
\newblock \showarticletitle{The division problem with single-peaked
  preferences: a characterization of the uniform allocation rule}.
\newblock \bibinfo{journal}{\emph{Econometrica}}  \bibinfo{volume}{59}
  (\bibinfo{year}{1991}), \bibinfo{pages}{509--519}.
\newblock


\end{thebibliography}

\newpage
% Appendix
\appendix

\section{Group Lottery}\label{appendix:GL}

\begin{algorithm}
\caption{\label{alg:simultaneous}}
\KwIn{A finite set $E$}
\KwOut{A random order \(\rOrder \in \o_E\)}
Choose $S\subset E$. Generate
\begin{enumerate}%[label=\roman*.]
\item[i.]a uniform random order $\rOrder^S \in \o_S$,
\item[ii.]a uniform random order $\rOrder^- \in \o_{E \backslash S}$.
\item[iii.] a uniform random subset $\positions \subset \{1, \ldots, |E|\}$ with $|\positions| = |S|$.
\end{enumerate}
Generate $\rOrder$ from $\rOrder^S, \rOrder^-, \positions$ placing elements of $S$ in positions $\positions$, maintaining the order of elements of $S$ as given by $\rOrder^S$ and the order of elements of $E \backslash S$ as given by $\rOrder^-$.
\end{algorithm}
\begin{lemma} \label{lem:uniform-random-order}
For any finite set \(E\), Algorithm~\ref{alg:simultaneous} generates a uniform random order $\rOrder \in \o_{E}$: for each order $\order \in \o_{E},$ $P(\rOrder=\order) = 1/|E|!$.
\end{lemma}

\proof[Proof of Lemma~\ref{lem:uniform-random-order}]
Fix an order $\order \in \o_{E}$. Let $\order_S$ and $\order_{-S}$ be the restriction of $\order$ to $S$ and $E \setminus S$, respectively. For any \(i \in E\), we let \(p_i\) be the position of \(i\) in \(\order\), that is, \(\order_{p_i} = i\). We define \(\positions_S = \cup_{i \in S} \{p_i\}\).  In order to end with the order $\order$, it must be that:

\begin{itemize}
     \item The order \(\rOrder\) generated in step i. is equal to $\order_{S}$, which occurs with probability $\bP(\rOrder=\order_S) = 1/|S|!$.
     \item The order \(\rOrder^{-}\) generated in step ii. is equal to $\order_{-S}$, which occurs with probability $\bP(\rOrder^{-}=\order_{-S}) = 1/(|\E|-|S|)!$.
     \item The random subset $\positions$ generated in step iii. is equal to \(\positions_S\), which occurs with probability $\bP(\positions=\positions_{S}) = \prod_{j=0}^{|S| - 1} (|S| - j)/(|E| - j)$.
 \end{itemize}
Hence, the probability that the algorithm generates the order $\order$ is
\[\frac{1}{|S|!}\frac{1}{E-|S|!}\prod_{j=0}^{|S| - 1} \frac{|S| - j}{|E| - j} = \frac{1}{|E|!}.\]
\endproof

\subsection{Incentives}
\proof[Proof of Proposition~\ref{prop:GL-incentives}]
Fix an arbitrary agent $i \in \sN$. Let $\a$ be an action profile such that $\a_{G_i}$ is the group request strategy and $\a_{-G_i}$ is arbitrary. Let $\a'$ denote an alternative strategy profile in which $a'_j = a_j$ for $j \not \in G_i$. Let $V$ be the set of valid groups according to $\a$, $V'$ the set of valid groups according to $\a'$, and $V^-$ be the set of valid groups not containing any members of $G_i$:
\begin{align*}
V \,\,& = \{S \subset \sN: a_j = S~ \forall j \in S\}.\\
V' & = \{S \subset \sN: a'_j = S~ \forall j \in S\} \\
V^- & = \{S \subset \sN\setminus G_j : a_j = S~ \forall j\in S\}.
\end{align*}
Note that $V^- \subseteq V \cap V'$, and that agents in $G_i$ do not influence $V^-$. We generate uniform random orders $\rOrder$ and $\rOrder'$ over $V$ and $V'$ (respectively) using Algorithm~\ref{alg:simultaneous}: we first generate a uniform random order $\rOrder^-$ over $V^-$, and then extend this to obtain $\rOrder$ and $\rOrder'$. We will prove that for any realization of $\rOrder^-$,
\begin{equation}\label{eq:GL:incentives}
\E[u_i(\xgl(\a,\rOrder))|\rOrder^-] \ge \E[u_i(\xgl(\a',\rOrder'))|\rOrder^-].
\end{equation}
Because agents in $G_i$ cannot influence $\rOrder^-$, it follows immediately that the unconditional expected utility of agent $i$ is also higher under the group request strategy.

If $\sum_{S \in V^{-}} |S| \le k - |G_i|$, then the group request strategy guarantees that all members of $G_i$ will receive a ticket, so there is nothing to prove. Otherwise, let
\[\tau(\rOrder^-) = \tau(k - |G_i| + 1, \rOrder^-),\]
be as defined in \eqref{eq:def:tau} where the size function is the cardinality of the valid group declared by the agent, that is, \(|\order_t| = |a_{\order_t}|\). Intuitively, \(\tau\) is the first point at which the number of remaining tickets would be less than $|G_i|$, when processing valid groups in $V^{-}$ according to order $\rOrder^-$.

Because agents in $G_i$ follow the group request strategy under $\a$, we have $V = V^- \cup G_i$. Members of $G_i$ get a payoff of 1 if and only if $G_i$ is in the first $\tau(\Sigma^-)$ positions of $\Sigma$. Therefore,
\begin{equation} \E[u_i(\xgl(\a,\Sigma))|\Sigma^-] = \frac{\tau(\Sigma^-)}{1+|V^-|}. \label{eq:ui-gr} \end{equation}

We now turn to the action profile $\a'$. Because the Group Lottery gives at most one ticket to each agent, $i$ gets a payoff of 1 if and only if all members of $G_i$ receive a ticket. This is not possible unless (i) every agent in $G_i$ is included in a valid group in $V'$, and (ii) in the order $\Sigma'$, all valid groups in $V' \backslash V^-$ appear before group $S = \Sigma^-_{\tau(\Sigma^-)}$. According to the algorithm, the conditional probability of (ii) given $\Sigma^-$ is at most
\[\frac{\tau(\Sigma^-)}{2 + |V^-|}\frac{\tau(\Sigma^-)-1}{1 + |V^-|},\]
which is smaller than the right side of~\eqref{eq:ui-gr}, implying that group $G_i$ has not benefited from its deviation.

Next, we show that any other strategy is not dominant. Let $j \not\in G_i$ and $\tilde \a$ denote an action profile such that $j \in \tilde a_i$, $i \not\in \tilde a_j$ and the remaining actions $\tilde\a_{-\{i,j\}}$ are arbitrary. Under $\tilde \a$ agent $i$ is not in a valid group then it is not award and group $G_i$ get a payoff of 0. This is strictly less than the payoff under a group request, which is greater than the probability of $G_i$ being the first group to be processed. Therefore, we can restrict to strategies $\hat\a$ where $\hat a_{i'} \subset G_i$ for any $i'\in G_i$. Furthermore, $G_i$ will have a positive expected payoff only if under $\hat\a$ its members  are divided into two or more valid groups. Let actions $\hat\a_{-G_i}$ be such that $\hat a_j = \sN \setminus G_i$ for any $j \in \sN \setminus G_i$, and $\hat V$ be the set of valid groups according to $\hat a$,
\[\hat V \,\, = \{S \subset \sN: \hat a_j = S~ \forall j \in S\}.\]
Observe that $|\hat V| \ge 3$. By assumption $n>k$, so $G_i$ will get a payoff of 1 if and only if valid group $\sN \setminus G_i$ is the last valid group to be processed. This event occurs with probability
\[\frac{(|\hat V|-1 )!}{|\hat V| !} = \frac{1}{|\hat V|}.\]
This is strictly smaller than $1/2$ the expected utility when $G_i$ select a group request.
\endproof
\subsection{Performance}
\begin{lemma}\label{lemma:gl-small-groups-advantage}
Fix any instance and any pair of agents \(i, j \in \sN\). Let \(\a\) be an action profile under the Group Lottery such that \(G_i\) and \(G_j\) select the group request strategy. If \(|G_i| \ge |G_j|\), then
\begin{equation}
u_i(\GL(\a)) \le u_j(\GL(\a)).
\end{equation}
\end{lemma}
\proof[Proof of Lemma~\ref{lemma:gl-small-groups-advantage}]
Let \(V\) be the set of valid groups given \(\a\). Observe that by assumption \(G_i, G_j \in V\). We define \(S_i, S_j\) to be the set of orders over \(V\) that guarantee a payoff of \(1\) to group \(G_i\) and \(G_j\), respectively. It suffices to show that \[|S_j| \ge |S_i|.\] To prove this, we will construct an injective map \(f:S_i\to S_j\). We let \(f\) be the map that only swap the positions of \(G_i\) and \(G_j\), keeping all the remaining positions unchanged. Clearly \(f\) is injective. Thus, all that remains to show is that for any \(\order \in S_i\), \(f(\order) \in S_j\). Fix \(\order \in S_i\). If \(\order \in S_j\), then there are enough tickets to satisfy both groups and \(f(\order) \in S_i \cap S_j \subseteq S_j\). If \(\order \not\in S_j\), then then the number of tickets remaining before \(G_j\) is processed under \(f(\order)\) is the same as the number of tickets remaining before \(G_i\) is processed under \(\order\). Because \(\order \in S_i\), this number is at least \(|G_i|\) which by assumption is at least \(|G_j|\), so $f(\order) \in S_j$.
\endproof
\begin{fact}\label{fact-ratio-increasing}
For any  \(a,b,c \in \R\) such that \(a \le b\) and \(c\ge 0\), then \begin{equation}\frac{a}{b} \le \frac{a+c}{b+c}.\end{equation}
\end{fact}
\proof[Proof of Proposition~\ref{lemma:mart-bounds}]
First, we will prove the right inequality of~\eqref{eq:mart-bounds}.  If \(k \ge \sum_i a_i - \bar a + 1\), then our upper bound is at least \(n\) and immediately holds. Hence, without loss of generality we can assume \(k \le \sum_i a_i - \bar a\). We let \(S_t = S_t(\rOrder)\) be the sum of the first \(t\) numbers according to  \(\rOrder\), that is, \begin{equation}
S_t = \sum_{i=1}^ta_{\rOrder_i}.
\end{equation}
We define
\begin{equation}
    Z_t^* = \frac{S_t - t \mu}{n-t}.
\end{equation}
As mentioned in~\citet{serfling1974probability}, the sequence \(Z^*_1,\ldots,Z^*_{n-1}\) is a forward martingale. Furthermore, \(k \le \sum_i a_i -\bar a\) implies \(\bP (\tau\le n-1) = 1\) so \(\tau\) is bounded and \(Z^*_{\tau}\) is well defined. Hence, we can apply Doob's optional stopping theorem to obtain
\begin{equation}\label{OST}
\E\left[\frac{S_\tau - \tau \mu}{n-\tau}\right]= \E[Z^*_{\tau}] = \E[Z^*_1]= 0.
\end{equation}
From the definition of \(\tau\), we get
\begin{equation}
\E\left[\frac{S_\tau}{n-\tau}\right] \le (k + \bar{a} - 1)\E\left[\frac{1}{n-\tau}\right] \label{partial-sum-ub-2}.
\end{equation}
We claim that
\begin{equation}\label{tau-ratio-ub-2}
\E[\tau] \E\left[\frac{1}{n-\tau}\right] \le \E\left[\frac{\tau}{n-\tau}\right].
\end{equation}
For any \(x < n\), define \[f(x) = \left(x-\E[\tau]\right)\left(\frac{1}{n-x}-\frac{1}{n-\E\left[\tau\right]}\right).\] Note that \(f(x) \ge 0\) for all \(x\), so \(\E[f(\tau)] \ge 0\).
Thus,
\[0 \leq \E[f(\tau)] = \E\left[(\tau-\E\left[\tau\right])\left(\frac{1}{n-\tau}-\frac{1}{n-\E\left[\tau\right]}\right)\right] = \E\left[\frac{\tau-\E\left[\tau\right]}{n-\tau}\right].\]
Combining equations~\eqref{OST},~\eqref{partial-sum-ub-2} and~\eqref{tau-ratio-ub-2} yields our desired result.

Now, we prove the left inequality of~\eqref{eq:mart-bounds}. If \(k \le \bar a\), then our lower bound is at most \(1\) and immediately holds, then  without loss of generality we can assume \(k \ge \bar a + 1\). To construct \(\rOrder\) we will generate a random order \(\rOrder'\) and iterate through it backwards, that is, \(\rOrder_t = \rOrder'_{n-t+1}\)  for \(t=1,\ldots,n\). We claim that for every \(\rOrder\),
\begin{equation}\label{taus-complement}
\tau(k,\Sigma) + \tau(\sum_i a_i - k + 1,\Sigma') = n+1.
\end{equation}
It suffices to show that
\begin{align}
\sum_{t=1}^{\tau(\sum_i a_i - k + 1,\Sigma')} a_{\rOrder'_t} &= \sum_{t=\tau(k,\Sigma)}^{n} a_{\rOrder_t}.
\end{align}
From the definition of \(\tau\), we have that
\[\sum_{t=\tau(k,\Sigma)+1}^{n} a_{\rOrder_t} = \sum_{i=1} a_i - \sum_{t=1}^{\tau(k,\Sigma)} a_{\rOrder_t} \le \sum_{i=1} a_i-k.\]
Similarly,
\[\sum_{t=\tau(k,\Sigma)}^{n} a_{\rOrder_t} = \sum_{i=1} a_i - \sum_{t=1}^{\tau(k,\Sigma)-1} a_{\rOrder_t} \ge \sum_{i=1} a_i-k+1.\]
Applying the upper bound in~\eqref{eq:mart-bounds} to \eqref{taus-complement} immediately implies
\[\E[\tau(k,\Sigma)] = n+1 - \E[\tau(\sum_i a_i - k + 1,\Sigma')] \geq n+1 - \frac{\sum_i a_i - k + \bar a}{\mu} = 1 + \frac{k-\bar a}{\mu}.\]
We now turn to equation~\eqref{tau-diff}. For any order \(\sigma\), we define \(h(\order)\) to be the order that: (i) is identical to \(\order\) from position \(\tau(k+k', \order)\) until the end, and (ii) flip the ordering of all elements from position \(1\) to \(\tau(k+k', \order) - 1\). More precisely,
\begin{equation}\label{def-h-map}
h(\order)_t = \begin{cases}
\order_t &\text{if }t \ge \tau(k + k', \order),\\
\order_{\tau(k + k', \order) - t} &\text{if }t < \tau(k + k', \order).
\end{cases}
\end{equation}
We claim that \[\tau(k + k', \order) = \tau(k + k', h(\order)).\] This implies that \(h(h(\order))) = \order\), which further implies that \(h\) is a bijective map. This follows from 2 observations: (i) the elements in the first \(\tau(k+k', \order) - 1\) positions are the same in both orders, and do not sum to \(k+k'\). Additionally, (ii) the agents in the first \(\tau(k+k', \order)\) positions are also the same in both orders, and they do sum to \(k+k'\).

We will show that for any \(\order \in \o_{[n]}\),
\begin{equation}\label{tau-split-ineq}
\tau(k, \order) + \tau(k', h(\order)) \ge \tau(k + k', \order).
\end{equation}
This implies our result as
\begin{align*}
\E[\tau(k, \rOrder) + \tau(k', h(\rOrder))]
&= \frac{1}{n!} \left(\sum_{\order \in \o_{[n]}}\tau(k, \order) + \tau(k', \order)\right)\\
&= \frac{1}{n!} \left(\sum_{\order \in \o_{[n]}}\tau(k, \order) + \tau(k', h(\order))\right)\\
&\ge \frac{1}{n!} \left(\sum_{\order \in \o_{[n]}}\tau(k + k', \order)\right)\\
&= \E[\tau(k + k', \rOrder)].
\end{align*}
The second equality follows as \(h\) is a bijective map, and the inequality follows from~\eqref{tau-split-ineq}. Thus all that remains is to show~\eqref{tau-split-ineq}. From the definition of \(\tau\), we have
\begin{align*}
\sum_{t=1}^{\tau(k, \order)} a_{\order_t} &\ge k,\\
\sum_{t=1}^{\tau(k + k', \order) - 1} a_{\order_t} &< k + k'.
\end{align*}
Implying that
\begin{equation}\label{tau-k'}
\sum_{t=\tau(k, \order) + 1}^{\tau(k + k', \order) - 1} a_{\order_t} < k'.
\end{equation}
Moreover, from the definition of \(h\) in~\eqref{def-h-map} it follows that
\begin{equation}\label{map-h-order}
\{h(\order)_{1},\ldots,h(\order)_{\tau(k + k', \order) - \tau(k, \order) - 1}\} = \{\order_{\tau(k + k', \order) - 1},\ldots,\order_{\tau(k, \order) + 1}\}.
\end{equation}
Combining~\eqref{tau-k'} and~\eqref{map-h-order} yields
\[\sum_{t=1}^{\tau(k + k', \order) - \tau(k, \order) - 1} a_{h(\order)_t} < k'.\]
This implies~\eqref{tau-split-ineq} as by definition \(\tau\) is integral and
\[\tau(k', h(\order)) > \tau(k + k', \order) - \tau(k, \order) - 1.\]
\endproof

\proof[Proof of Lemma~\ref{lemma:gl-utility-bounds}]
Because all groups are playing the group request, the set of valid groups is \(\G\). In what follows we fix an arbitrary agent \(i \in \sN\). We construct a random order \(\rOrder\) over \(\G\) using Algorithm~\ref{alg:simultaneous}: we generate an order \(\rOrder^{-i}\) over \(\G\setminus G_i\) and then extend it to \(\G\). By Lemma~\ref{lem:uniform-random-order}, the resulting order $\rOrder$ is uniformly distributed. We let \(\tau^{-i} = \tau(k - |G_i| + 1,\rOrder^{-i})\) be the number of positions in \(\rOrder\) that ensure \(G_i\) a payoff of 1 given \(\rOrder^{-i}\). Note that \(\tau^{-i}\) is well defined as \(k < n\) implies \(k - |G_i| + 1 \le n - |G_i| = \sum_t |\rOrder^{-i}_t|\). Moreover, if \(G_i\) is in the first \(\tau^{-i}\) positions of \(\rOrder\), then it gets a payoff of 1 as the number of remaining tickets before it is processed is at least
\[k - \sum_{t=1}^{\tau^{-i}-1}|\rOrder^{-i}_t| \ge k - (k-|G_i|) = |G_i|.\]
On the other hand, if \(G_i\) is in the last \(m - \tau^{-i}\) positions of \(\rOrder\), then it gets a payoff of \(0\) because the number of remaining tickets when $G_i$ is processed is at most
\[k - \sum_{t=1}^{\tau^{-i}}|\rOrder^{-i}_t| \le k - (k-|G_i|+1) = |G_i| - 1.\]
Therefore,
\begin{equation}\label{eq:conditional-uti-Gi}
u_i(\xgl(\a,\rOrder)) = \E[\E[u_i(\xgl(\a,\rOrder))\vert \rOrder^{-i}]] = \frac{\E[\tau^{-i}]}{m}.
\end{equation}
By Proposition~\ref{lemma:mart-bounds} equation~\eqref{eq:mart-bounds}, we have
\begin{align}
\E[\tau^{-i}] \le \frac{k- G_i + \max_G |G|}{(n-|G_i|)/(m-1)}.
\end{align}
This and equation~\eqref{eq:conditional-uti-Gi} yields
\begin{align}\label{GL-utility-almost-bounds}
u_i(\GL(\a)) \le \left(\frac{k- G_i + \max_G |G|}{n-|G_i|}\right)\left(\frac{m-1}{m}\right).
\end{align}
Because there is a group of size \(|G_i|\) and the remaining \(n-|G_i|\) agents can be in at most \(n-|G_i|\) groups of size \(1\), we have
\begin{equation}
m \leq n - |G_i| + 1.
\end{equation}
This implies
\begin{equation}\label{GL-m-ratio}
\frac{m-1}{m} \leq \frac{n - |G_i|}{n - |G_i|+1}.
\end{equation}
From~\eqref{GL-utility-almost-bounds} and~\eqref{GL-m-ratio} it follows that
\begin{equation}
u_i(\GL(\a)) \le \frac{k- G_i + \max_G |G|}{n-|G_i| + 1}.
\end{equation}
Applying Fact~\ref{fact-ratio-increasing} with \(a = k + \max_G|G| - |G_i|\), \(b = n  + 1 - |G_i|\) and \(c=|G_i| - 1\), we get
\begin{equation}\label{gl-utility-ub-2}
\frac{k + \max_G|G| - |G_i|}{n - |G_i| + 1} \le \frac{k + \max_G|G| - 1}{n} \le \frac{k}{n} \left(1 + \kappa\right).
\end{equation}
Note that to apply Fact~\ref{fact-ratio-increasing} we need \(a \le b\), we can assume this without loss of generality. Otherwise, \(a > b\) or equivalently
\begin{equation}\label{assumption-k-s-n}
k + \max_G|G| > n  + 1.
\end{equation}
We claim that if the inequality above holds then \(\frac{k}{n} \left(1 + \kappa\right) > 1\), hence, our bound is vacuous. This can be seen by noting that
\begin{equation}
1 + \kappa \ge \frac{k + \max_G|G| - 1}{k} > \frac{n}{k}.
\end{equation}
The last inequality follows by~\eqref{assumption-k-s-n}.
\endproof

\proof[Proof of Theorem~\ref{thm:gl-is-good}]
Fix \(\alpha,\kappa \in (0,1)\) and an arbitrary instance in \(I(\kappa, \alpha)\). In what follows, we let \(\a\) denote the group request action profile under the Group Lottery. We define \(s\) to be the maximum group size, that is, \(s=\max_{G \in \G}|G|\).

We begin with the efficiency guarantee.  For any order $\order \in \o_\G$,
\begin{equation} U(\xgl(\a,\order)) \ge 1 - \frac{s-1}{k} \geq 1 - \kappa, \label{eq:gl-efficiency}\end{equation}
where the second inequality follows as our instance is in \(I(\kappa, \alpha)\). This is fairly trivial: if we let
\[\tau(\order)= \tau(k+1, \order),\]
be as defined in \eqref{eq:def:tau} where the size of a group is its number of elements. Then $U(\xgl(\a,\order))$ is exactly $\frac{1}{k} \sum_{j = 1}^{\tau(\order,k+1)-1}|\order_j|$, which is at least $\frac{1}{k}(k - (s-1))$, because adding one more group (of size at most $s$) brings the sum above $k$. From~\eqref{eq:gl-efficiency}, it immediately follows that if $\rOrder$ is a random order on $\G$, then
\[\E[U(\xgl(\a,\rOrder))] \ge 1 - \kappa.\]

We now show that in this setting the outcome is \((1 - 2\kappa)\)-fair. Our goal is to show that for any pair of agents \(i,j \in \sN\),
\begin{equation}
\frac{u_i(\GL(\a))}{u_j(\GL(\a))} \ge 1-2\kappa.
\end{equation}
By Lemma \ref{lemma:gl-small-groups-advantage}, we can assume without loss of generality that
\begin{align}
|G_i| &= \max_{G\in \G} |G| \text{ and } |G_j| =  \min_{G\in\G}|G|.
\end{align}
We let \(\mu^{-i}\) be the average group size in \(\G\setminus G_i\), more precisely,
\begin{equation}\label{defs-Gi}
\mu^{-i} = \frac{n-|G_i|}{m-1}.
\end{equation}
We claim that
\begin{align}\label{GL-fairness-ratio}
\frac{u_i(\GL(\a))}{u_j(\GL(\a))} &\ge \frac{k - 2|G_i| + 1 + \mu^{-i}}{k - |G_j|  + \mu^{-i}}.
\end{align}
This implies our result as
\[\frac{k - 2|G_i| + 1 + \mu^{-i} }{k - |G_j|  + \mu^{-i}} \ge \frac{k - 2|G_i| + 1 + |G_j|}{k} \ge 1 - \frac{2(|G_i| - 1)}{k} \ge 1 -2\kappa.\]
In the first inequality, we apply Fact~\ref{fact-ratio-increasing} with \(a = k - 2|G_i| + 1 + |G_j|\), \(b = k\) and \(c = \mu^{-i} - |G_j|\). The last inequality follows from the definition of \(I(\kappa,\alpha)\) in~\eqref{alpha-kappa-family}. Remember that for any instance in \(I(\kappa, \alpha)\), we have
\[\frac{|G_i|-1}{k} = \frac{\max_{G\in\G} |G|-1}{k} \leq \kappa.\]
We now turn to the proof of equation~\eqref{GL-fairness-ratio}. Let \(\rOrder^{-i}\) be a uniform order on \(\G\setminus G_i\). Applying Lemma~\ref{lemma:gl-utility-bounds} to agents \(i,j\), it follows that
\begin{equation}
\frac{u_i(\xgl(\a,\rOrder))}{u_j(\xgl(\a,\rOrder))}
= \frac{\E[\tau(k-|G_i|+1,\rOrder^{-i})]}{\E[\tau(k-|G_j|+1,\rOrder^{-j})]}.
\end{equation}
By Proposition~\ref{lemma:mart-bounds} equation~\eqref{tau-diff}, we have
\begin{equation}\label{tau-split}
\E[\tau(k-|G_j|+1,\rOrder^{-j})] \le \E[\tau(k-|G_i|+1,\rOrder^{-j})] + \E[\tau(|G_i|-|G_j|,\rOrder^{-j})].
\end{equation}
We claim that for any constant \(c \in \N\), such that \(c\le n-|G_i| = \sum_t |\rOrder^{-i}_t|\),
\begin{equation} \E[\tau(c,\rOrder^{-j})] \le \E[\tau(c,\rOrder^{-i})]. \label{eq:coupling} \end{equation}
We now show~\eqref{eq:coupling}. We generate \(\rOrder^{-i}\) using  \(\rOrder^{-j}\) in the following way:
\[\rOrder^{-i}_t(\rOrder^{-j}) = \begin{cases} \rOrder^{-j}_t &\text{if }\rOrder^{-j}_t \ne G_i,\\
G_j &\text{otherwise}.
\end{cases}\]
Note that by construction for any \(\rOrder^{-j}\), \(\tau(c, \rOrder^{-i}(\rOrder^{-j})) \ge \tau(c, \rOrder^{-j})\). This establishes~\eqref{eq:coupling}. Applying~\eqref{eq:coupling} twice we get
\begin{equation}\label{taus-domination}
\E[\tau(k-|G_i|+1,\rOrder^{-j})] + \E[\tau(|G_i|-|G_j|,\rOrder^{-j})]
\le \E[\tau(k-|G_i|+1,\rOrder^{-i})] + \E[\tau(|G_i|-|G_j|,\rOrder^{-i})].
\end{equation}
Then by~\eqref{tau-split} and~\eqref{taus-domination},
\begin{equation}
\frac{\E[\tau(k-|G_i|+1,\rOrder^{-i})]}{\E[\tau(k-|G_j|+1,\rOrder^{-j})]} \ge \frac{\E[\tau(k-|G_i|+1,\rOrder^{-i})]}{\E[\tau(k-|G_i|+1,\rOrder^{-i})] + \E[\tau(|G_i|-|G_j|,\rOrder^{-i})]}.
\end{equation}
By Proposition~\ref{lemma:mart-bounds} equation~\eqref{eq:mart-bounds}, we have
\begin{align}
\E[\tau(k-|G_i|+1,\rOrder^{-i})] &\ge \frac{k - 2|G_i| + 1 + \mu^{-i}}{\mu^{-i}},\label{GL-tau-ub}\\
\E[\tau(|G_i|-|G_j|,\rOrder^{-i})] &\le \frac{2|G_i|-|G_j|-1}{\mu^{-i}}.\label{GL-tau-lb}
\end{align}
From~\eqref{GL-tau-ub} and~\eqref{GL-tau-lb}, we have
\begin{align*}
\frac{\E[\tau(k-|G_i|+1,\rOrder^{-i})]}{\E[\tau(k-|G_i|+1,\rOrder^{-i})] + \E[\tau(|G_i|-|G_j|,\rOrder^{-i})]}
&\ge \frac{k - 2|G_i| + 1 + \mu^{-i}}{k - 2|G_i| + 1 + \mu^{-i} + 2|G_i|-|G_j|-1}\\
&= \frac{k - 2|G_i| + 1 + \mu^{-i}}{k - |G_j| + \mu^{-i}}.
\end{align*}
\endproof

\subsubsection{Tightness}

\begin{proposition}\label{prop:gl-tightness}
For any \(\epsilon>0\) and $\alpha \in (0,1)$, there exists \(\kappa \in (0,1)\) and an instance in \(I(\kappa, \alpha)\) such that the group request equilibrium outcome of the Group Lottery is not \(1 - (2-\epsilon) \kappa\) fair.
\end{proposition}

\proof[Proof of Proposition~\ref{prop:gl-tightness}]
We consider an instances in which there is one ``single person" (group of size one), and the remaining groups are ``couples" (groups of size two). Fix $r \in \mathbb{N}$ such that
\begin{equation} \frac{1}{r} < \epsilon, \label{eq:r-choice} \end{equation}
and set $k = 2r-1$ and $\kappa = 1/k$. Let $m$ be the total number of groups, and note that for all sufficiently large $m$, the instance will be in $I(\kappa,\alpha)$.

The single person is successful if among the first $r$ groups (which occurs with probability $\frac{r}{m}$), while each couple is successful if among the first $r-1$ couples (which occurs with probability $\frac{r-1}{m-1}$). Thus, as $m \rightarrow \infty$, the ratio of the utility each couple to the utility of the single person converges to
\[\frac{r-1}{r} < 1 - \frac{2-\epsilon}{2r-1} = 1 - (2-\epsilon) \kappa ,\]
where the first equality follows from \eqref{eq:r-choice}.
\endproof

\subsection{Proof of Proposition~\ref{prop:badnews}}\label{appendix:RA}

We divide the proof into two parts: Lemma~\ref{lemma:performance-lb} establishes that for any instance in \(I(\kappa,\alpha)\), there exists a random allocation that is $(1-\kappa)$-efficient and fair. Lemma~\ref{prop:GL-optimal} shows that improving beyond $(1-\kappa)$-efficiency may require abandoning even approximate fairness.

\begin{lemma}\label{lemma:performance-lb}
Fix $\kappa,\alpha \in (0,1)$. For every instance in $I(\kappa,\alpha)$, there exist a random allocation that is $(1-\kappa)$-efficient and fair.
\end{lemma}

\proof[Proof of Lemma~\ref{lemma:performance-lb}]
We claim that there exists of a random allocation \(\pi^*\) such that for every agent \(i \in \sN\),
\begin{equation}\label{eq:utility-opt-random-allo}
u_i(\pi^*) = u^* = \frac{k - \max_{G \in \G}|G| + 1}{n}.
\end{equation}
It immediately follows that \(\pi^*\) is fair as the utility of every agent is equal to \(u^*\). Moreover, the utilization in this system is
\[U(\pi^*) =\frac{nu^*}{k}  = 1 - \frac{\max_G |G| - 1}{k} \ge 1 - \kappa.\]
The inequality follows because our instance is in \(I(\kappa,\alpha)\).

All that remains is to prove our claim. To this end, we will apply Theorem 2.1 in \citet{nguyen2016assignment} which establishes that any utility vector such that (i) the sum of all agents' utilities is at most \(k - \max_{G \in \G}|G| + 1\), and (ii) members of each group receive the same utility, can be induced by a lottery over feasible allocations. Before formally presenting this result some definitions are needed.

A {\em group allocation} is represented by \(x \in \{0,\ldots,k\}^m\) satisfying \(\sum_{G\in\G}x_G \le k\), where \(x_G\) represents the number of tickers assigned to group \(G\). For simplicity, we restrict to allocations such that for every group \(G\), \(x_G \in \{0,|G|\}\). Notice that a group \(G\) is successful if and only if \(x_G = |G|\).   A {\em random group allocation} correspond to a distribution \(\pi\) over the set of group allocations. A {\em group utility vector} \(u \in [0,1]^m\) associates to each group \(G\) a utility \(u_G\).

Theorem 2.1 in \citet{nguyen2016assignment} establishes that if a group utility vector \(u'\) satisfies
\begin{equation}\label{eq:nguyen-condition}
\sum_{G\in\G}|G|u'_G \le  k - \max_{G \in \G}|G| + 1,
\end{equation}
then it can be induced by a random group allocation, that is, there exists a random allocation \(\pi\) such that
\[u_i(\pi) = u'_G \text{ for every } G \in \G, i \in G.\]
Observe that the expected number of tickets awarded to each group is equal to the sum of the utilities of its members. Therefore, condition~\eqref{eq:nguyen-condition} can be interpreted as an upper bound on the total expected number of tickets awarded, which depends on the maximum number of tickets demanded by a single group.

Hence, the existence of a random allocation \(\pi^*\) that yields~\eqref{eq:utility-opt-random-allo} follows by letting \(\hat{u}\) be the utility vector that gives each group a utility of \(u^*\). Note that \(\hat{u}\) satisfies condition~\eqref{eq:nguyen-condition} as
\[\sum_{i\in\sN}\hat{u}_i =  nu^* = k - \max_{G \in \G}|G| + 1.\]
\endproof

\begin{lemma}\label{prop:GL-optimal}
For any $\epsilon > 0$, there exists \(\alpha,\kappa \in (0,1)\) and an instance in \(I(\kappa, \alpha)\) such that no random allocation is $\epsilon$-fair and $(1 - \kappa + \epsilon)$-efficient.
\end{lemma}
\proof[Proof of Lemma~\ref{prop:GL-optimal}]
Fix $\epsilon > 0$ and let \(m,s,r \in \N\). We consider an instance with \(k = rs - 1\) tickets, one group of size \(s-1\) and \(m-1\) groups of size \(s\). Let \(i\) be a member of the group of size \(s-1\).  Let \(\ra\) be any random allocation. Because at most \(r-1\) of the large groups can be satisfied in any deterministic allocation, we have that
\begin{equation}\label{eq:GL-optimal-ub}
s(m-1)\min_{j \not\in G_i}\{u_j(\ra)\} \le \sum_{j \not\in G_i} u_j(\ra) \leq s(r-1).\end{equation}
If the allocation is $\epsilon$-fair, then it must be the case that
\begin{equation} \epsilon u_i(\ra) \leq \min_{j \not\in G_i} \{u_j(\ra)\} \leq \frac{r-1}{m-1},\end{equation}
where the right inequality follows from~\eqref{eq:GL-optimal-ub}.

Observe that if \(G_i\) is successful then there will be no tickets wasted; otherwise, there will be \(s-1\) tickets wasted. Hence, the utilization under \(\ra\) is
\begin{equation}\label{eq:GL-optimal-utilization}
1 - \left(1 - u_i(\ra)\right) \frac{s - 1}{k} \le 1 - \left(1 - \frac{r-1}{\epsilon(m-1)}\right) \frac{s - 1}{k}.
\end{equation}

If we choose \(r,m\) such that $(r-1)/(m-1) < \epsilon^2 k/(s-1)$, and define $\kappa = (s-1)/k$ and $\alpha = k/n$, then our instance is in \(I(\kappa,\alpha)\), and \eqref{eq:GL-optimal-utilization} implies that utilization is strictly smaller than \(1 -\kappa + \epsilon\).
\endproof

\newpage
\section{Individual Lottery}\label{appendix:IL}

\subsection{Incentives}

\proof[Proof of Proposition~\ref{prop:IL-incentives}]
This is a direct consequence of Proposition~\ref{prop:IL(ell)-incentives}. Notice that for any given instance with \(k\) tickets, the Individual Lottery is equivalent to the Individual Lottery with limit \(\ell = k\). Therefore, \(r = \lceil |G|/k\rceil=1\) and our result follows.
\endproof

\proof[Proof of Proposition~\ref{prop:IL-monotonicity}]
Consider any agent \(i\). We let \(\a_{-i} \in A_{-i}\) be an arbitrary set of actions and \(a_i' > a_i \ge |G_i|.\) We begin by showing that for any order over agents \(\order\), the conditional expected utility of \(i\) is the same under both strategies. We consider two possible cases. First, the number of tickets remaining before agent \(i\) is processed is \(a_i\) or less. Then, under both strategies the allocation of every agent is the same and the payoff of \(G_i\) coincide. Second, the number of tickets remaining before agent \(i\) is processed is greater than \(a_i\). Then, under both strategies agent \(i\) obtains at least \(|G_i|\) tickets and the group gets a payoff of \(1\).

Now, we will show that the utility of every group \(G\ne G_i\) is weakly better under \(a_i\). It suffices to show that for any order \(\sigma \in \o_N\),
\begin{equation}\label{eq:IL-monotonicity}
\xil_j((a_i,\a_{-i}),\sigma) \ge \xil_j((a_i',\a_{-i}),\sigma)~\text{for every agent }j\ne i.
\end{equation}
Because this holds for any order \(\order\), and the random order over agents used in the Individual Lottery is uniformly distributed, this implies our result. Let \(T = T(\sigma)\) be the position of agent \(i\) in \(\order\), that is, \(T = \{t \in \{1,\ldots,n\}: \sigma_t = i\}\). The allocation of agents \(\sigma_1,\ldots,\sigma_{T-1}\) is not affected by the action of \(i\), then~\eqref{eq:IL-monotonicity} holds. A smaller request can only lead to a smaller allocation, hence the allocation of agent \(i\) is weakly smaller under \(a_i\). Therefore, the allocation of agents \(\sigma_{T + 1}, \ldots, \sigma_n\) is weakly greater under \(a_i\) as the only difference is due to agent \(i\).
\endproof

\subsection{Performance}

\proof[Proof of Theorem~\ref{thm:il-is-bad}]
Consider an instance with \(n = rs\) agents divided into one large group of size \(s\) and \(s(r-1)\) small groups of size one. Besides, the number of tickets is \(k = \lfloor \alpha r \rfloor s\). Observe that for any \(s,r \in \N\),
\[\frac{k}{n} = \frac{\lfloor \alpha r \rfloor s}{rs} \le \alpha.\]
Thus, if \((s-1)/k \le \kappa \) then this instance will be in \(I(\kappa,\alpha)\). We claim that a sufficient condition for this is
\begin{equation}\label{eq:IL-thm-cond}
r\ge \frac{\kappa + 1}{\alpha\kappa}.
\end{equation}
This can be seen as
\[\frac{s-1}{k} = \frac{s-1}{\lfloor \alpha r \rfloor s} \le \frac{1}{\alpha r - 1} \le \kappa.\]
In the first inequality we use that for any \(x\), \(\lfloor x \rfloor \ge x-1\) and \((s-1)/s \le 1\). The last inequality follows from~\eqref{eq:IL-thm-cond}. For ease of exposition, in the proof sketch of Theorem~\ref{thm:il-is-bad} we consider \(n\) agents, a large group of size \(s=n^{3/4}\) and \(k = \alpha n\) tickets.

Let agents \(i,j\) be such that \(|G_i| = 1\) and \(|G_j| = s\). We claim that for any dominant strategy equilibrium \(\a\),
\begin{equation}\label{eq:ub-utility-small-group-gr}
u_i(\IL(\a)) \le \frac{k}{s^2} \leq \frac{\alpha r}{s}.
\end{equation}
This bound is key to prove both guarantees. We start by proving the efficiency result. The expected utilization in this system is:
\begin{align*}
\frac{1}{k} \sum_{i' \in \sN}u_{i'}(\IL(\a)) &= \frac{s}{k}\left((r-1) u_i(\IL(\a)) + u_j(\IL(\a)\right)\\
&\le \frac{r-1}{s} + \frac{s}{k}\\
&= \frac{r-1}{s} + \frac{1}{\lfloor\alpha r\rfloor}.
\end{align*}
In the inequality we use that \(u_j(\IL(\a)) \le 1\) and the first inequality in~\eqref{eq:ub-utility-small-group-gr}. Hence, if we choose \(s=s(r)\) such that \(r/s(r)\to 0\) as \(r\) grows, then the right side goes to \(0\) as we take the limit.

We now turn to the fairness guarantee. Because the first agent to be processed always get a payoff of \(1\), we get that
\[u_j(\IL(\a)) \ge \frac{s}{n} = \frac{1}{r}.\]
Note that this lower bound is independent of \(s\), and is tight when all agents in small groups request \(k\) tickets.

Using this and the second inequality in~\eqref{eq:ub-utility-small-group-gr}, we obtain
\[\frac{u_i(\IL(\a))}{u_j(\IL(\a))} \le \frac{\alpha r^2}{s}.\]
Therefore, if we choose \(s=s(r)\) such that \(r^2/s(r)\to 0\) as \(r\) grows, then again the right side goes to \(0\) as we take the limit.

All that remains is to prove~\eqref{eq:ub-utility-small-group-gr}. We let \(\rOrder\) be a random order over agents. To generate \(\rOrder\) we use Algorithm~\ref{alg:simultaneous} from Lemma~\ref{lem:uniform-random-order}: set $S = G_{i} \cup G_{j}$, and independently generate (i) a uniform random order \(\rOrder^S\) over $S$, (ii) a uniform random order \(\rOrder^-\) over \(\sN \setminus S\), and (iii) uniform random positions $\positions \subseteq \{1, \ldots, |\sN|\}$ where agents in $S$ will be placed. By Lemma~\ref{lem:uniform-random-order}, the resulting order $\rOrder$ is uniformly distributed. Note that \(i\) will get a payoff 0 unless it appears in the first \(k/s\) positions of \(\rOrder^S\). Because \(\rOrder^S\) is uniformly distributed this event occurs with probability
\[\frac{k/s}{s+1} \le \frac{k}{s^2}.\]
This implies~\eqref{eq:ub-utility-small-group-gr} and concludes our proof.
\endproof

\subsection{Extension}

\begin{proposition}\label{prop:IL(ell)-incentives}
In the Individual Lottery with limit \(\ell\), the set of actions \(\a_{G}\) is dominant for group \(G\) if and only if \(\sum_{i \in S}a_i \geq |G|\) for all \(S \subseteq G\) such that \(|S| = \lceil |G|/\ell\rceil\).
\end{proposition}
\proof[Proof of Proposition~\ref{prop:IL(ell)-incentives}]
Fix an arbitrary agent $i$. Let \(\a_{-G_i} \in A_{-G_i}\) be an arbitrary action profile for agents not in $G_i$. We let \(r = \lceil |G_i|/\ell\rceil\) be the minimum number of members of \(G_i\) that must be awarded in the Individual Lottery with limit \(\ell\) in order for \(G_i\) to get a payoff of \(1\). Let \(\a_{G_i} \in A_{G_i}\) be any action profile such that \(\sum_{j \in S}a_j \geq |G_i|\) for all \(S \subseteq G_i\) such that \(|S| = r\).

First, we show that for any order \(\sigma \in \o_{\sN}\) the utility of agent \(i\) is maximized under \(\a_{G_i}\), that is,
\begin{equation}\label{eq:IL(ell)-incentives-1}
u_i(\xil((\a_{G_i},\a_{-{G_i}}),\sigma)) \ge u_i(\xil((\a'_{G_i},\a_{-G_i}),\sigma)), \text{ for every }\a'_{G_i} \in A_{G_i}.
\end{equation}
Because this holds for any order \(\sigma\), the expected utility of group \(G_i\) will also be maximized by \(\a_{G_i}\). Let \(T = T(\sigma)\) be the position of the \(r^{th}\) member of \(G_i\) under \(\sigma\):
\[T(\sigma) = \min\{t \in \{1,\ldots,n\}: \left| \sigma_{[t]} \cap G_i \right| = r\}.\]

If \(\sum_{j \in \sigma_{[T]}\setminus G_i} a_j > k - |G_i|\), then for any action profile selected by group \(G_i\) its payoff is 0.

If \(\sum_{j \in \sigma_{[T]}\setminus G_i} a_j \leq k - |G_i|\), then under action profile \(\a_{G_i}\) group \(G_i\) receives a payoff of 1.

Hence, \(\a_{G_i}\) maximizes the payoff of agent \(i\) for each \(\sigma\).

Second, let \(\hat \a_{G_i}\) be such that \(\sum_{j \in S}\hat a_j < |G_i|\) for some \(S \subseteq G_i\) with \(|S| = r\). We will show that there exists an order \(\hat\sigma \in \o_{\sN}\) such that
\[1 = u_i(\xil((\a_{G_i},\a_{-G_i}),\hat\sigma)) > u_i(\xil((\hat\a_{G_i},\a_{-G_i}),\hat\sigma)) = 0.\]
This combined with~\eqref{eq:IL(ell)-incentives-1} implies that \(\hat\a_{G_i}\) is dominated by  \(\a_{G_i}\). We construct \(\hat\sigma \) in the following way:
\begin{itemize}
     \item Agents in \(S\) are arbitrary placed in the first \(r\) positions of \(\hat\sigma\).
     \item Agents in \(\sN \setminus G_i\) are arbitrary placed in positions \(r+1,\ldots,r + n-|G_i|\) of \(\hat\sigma\).
     \item Agents in \(G_i\setminus S\) are arbitrary placed in the last \(|G_i|-r\) positions of \(\hat\sigma\).
 \end{itemize}
We begin by proving that if \(\a_{G_i}\) is selected then \(G_i\) received at least \(|G_i|\) tickets, implying that \(u_i(\xil((\a_{G_i},\a_{-G_i}),\hat\sigma))=1\). To see this note that the number of tickets received by \(G_i\) is
\[\sum_{j\in G}\xil_j((\a_{G_i},\a_{-G_i}),\hat\sigma) \ge \sum_{j\in S}\xil_j((\a_{G_i},\a_{-G_i}),\hat\sigma)\ge \min\{k,\sum_{j\in S}a_j\} \ge |G_i|.\]
The last inequality follows as \(k\ge|G_i|\) and \(\sum_{j\in S}a_j \ge |G_i|\). On the other hand, we show that when \(\hat \a_{G_i}\) is selected then \(G_i\) received strictly less than \(|G_i|\) tickets and \(u_i(\xil((\hat\a_{G_i},\a_{-G_i}),\hat\sigma))=0\). For the sake of contradiction, suppose that \(|G_i|\) received at least \(|G_i|\) tickets, then
\[\sum_{j\in \sN}\xil_j((\a_{G_i},\hat\a_{-G_i}),\hat\sigma) =
 \sum_{j\in G_i}\xil_j((\a_{G_i},\hat\a_{-G_i}),\hat\sigma) + \sum_{j\in \sN \setminus G_i}\xil_j((\a_{G_i},\hat\a_{-G_i}),\hat\sigma) \ge |G_i| + \sum_{j\in \sN \setminus G_i}1 = n.\]
 A contradiction, as \(k<n\). Note that \(G_i\) will get \(|G_i|\) or more tickets only if agent in position \(r+n-|G_i|+1\) is awarded, this implies that all agents in the first \(r+n-|G_i|\) must also be awarded.
\endproof

\proof[Proof of Proposition~\ref{prop:il-ell-bad}]
Consider a sequence of instances with \(n \rightarrow \infty\) and a constant \(k\) number of tickets. In each instance, there is one group of size \(1\) and the remaining groups have size \(\ell + 1\). We let \(i\) be such that \(|G_i| = 1\) and \(j\) be such that \(|G_j| = \ell + 1\). We let \(\rOrder\) be a uniform order over agents.

First, note that regardless of the action profile \(\a\), \(i\) gets utility \(1\) if among the first \(\lceil k/\ell\rceil\) agents in \(\rOrder\), and gets utility \(0\) if after the first \(k\) agents. Because \(\rOrder\) is drawn uniformly at random from \(\o_{\sN}\), we have
\begin{equation}\label{eq:IL(l)-1}
\frac{\lceil k/\ell\rceil}{n} \le u_{i}(\IL(\a)) \le\frac{k}{n}.
\end{equation}

Second, because \(\ell < |G_{j}|\), at least two agents from group \(G_{j}\) must be awarded in order for the group to get utility \(1\). Furthermore, any agent not among the first \(k\) will certainly not receive any tickets. Therefore, group \(G_{j}\) gets utility \(1\) only if some pair of agents from \(G_{j}\) are both among the first \(k\) agents. For any pair of agents, the chance that both are among the first \(k\) agents is $\binom{n-2}{k-2}/\binom{n}{k}$. Applying a union bound, we see that
\begin{equation}\label{eq:IL(l)-3}
u_{j}(\IL(\a)) \le  \frac{\binom{n-2}{k-2}\binom{\ell + 1}{2}}{\binom{n}{k}}=\frac{k(k-1)}{n(n-1)}\binom{\ell + 1}{2}.
\end{equation}
Combining the upper bounds derived in~\eqref{eq:IL(l)-1} and~\eqref{eq:IL(l)-3}, we bound the overall efficiency as follow
\begin{align*}
\frac{1}{k}\sum_{i' \in \sN} u_{i'}(\IL(\a))
&\leq  \frac{1}{k}\left(\frac{k}{n} + (n-1)\frac{k(k-1)}{n(n-1)}\binom{\ell+1}{2} \right)
 = \frac{2 +  (k-1)(\ell+1)\ell}{2n},
\end{align*}
which approaches zero as $n$ grows.

Furthermore, \eqref{eq:IL(l)-1} and \eqref{eq:IL(l)-3} imply that
\[
\frac{u_{j}(\IL(\a))}{u_{i}(\IL(\a))}\le \frac{\frac{k(k-1)}{n(n-1)}\binom{\ell + 1}{2}}{\frac{\lceil k/\ell\rceil}{n}} = \frac{k(k-1)}{\lceil k/\ell\rceil(n-1)}\binom{\ell+1}{2},\]
which also approaches zero as $n$ grows.
\endproof
\begin{proposition}\label{prop:il(ell)-eff}
For any \(\ell \in \N\) and any instance such \(\max_{G\in\G} |G| \le \ell \), every dominant strategy equilibrium outcome of the Individual Lottery with limit \(\ell\) is \(1/\ell-\)efficient.
\end{proposition}
\proof[Proof of Proposition~\ref{prop:il(ell)-eff}]
We let \(\rOrder\) be a uniform order over agents. We claim that if \(\a_{G_i}\) is dominant for \(G_i\), then
\begin{equation} \E[u_i(\xil(\a,\Sigma))]\ge \frac{k}{\ell n}. \label{eq:il-u-lb}\end{equation}
From this, it follows that if \(\a\) is such that all agents follow a dominant strategy, then
\[\E[U(\xil(\a,\Sigma))] = \frac{1}{k} \sum_{i \in \sN} \E[ u_i(\xil(\a,\Sigma))] \geq \frac{1}{\ell}.\]
We now prove \eqref{eq:il-u-lb}. If $|G_i| = 1$, then no matter the reports of others, $i$ succeeds if in the first $\lceil k/\ell \rceil$ positions, which occurs with probability $\frac{\lceil k/\ell \rceil}{n} \geq \frac{k}{\ell n}$.

Otherwise, because \(\max_{G\in\G} |G| \le \ell \) and agents in $G_i$ follow a dominant strategy, $i$ succeeds if any agent from $G_i$ is in the first $\lfloor k/\ell \rfloor$ positions.

If $k/\ell < 2$, then this occurs with probability $\frac{|G_i|}{n} \geq \frac{k}{\ell n}$. Thus, we turn to the case with $\min(|G_i|, k/\ell) \geq 2$. Fix two agents in $G_i$. The chance that at least one of them is in the first $\lfloor k/\ell \rfloor$ positions is
\begin{align} \frac{2\lfloor k/\ell \rfloor}{n} -  \frac{ \binom{n-2}{\lfloor k/\ell \rfloor-2} }{\binom{n}{\lfloor k/\ell \rfloor}} & \geq \frac{2\lfloor k/\ell \rfloor}{n} - \left(\frac{\lfloor k/\ell \rfloor}{n}\right)^2 \nonumber \\
& \geq \frac{k/\ell - 1}{n} + \frac{\lfloor k/\ell \rfloor}{n} - \left(\frac{\lfloor k/\ell \rfloor}{n}\right)^2 \nonumber \\
& = \frac{k}{\ell n} - \frac{1}{n} + \frac{\lfloor k/\ell \rfloor}{n}  \left(1 - \frac{\lfloor k/\ell \rfloor}{n}  \right). \nonumber
\end{align}
All that remains is to establish that
\[\frac{\lfloor k/\ell \rfloor}{n}  \left(1 - \frac{\lfloor k/\ell \rfloor}{n}  \right) \geq \frac{1}{n}.  \]
This holds because $\lfloor k/\ell \rfloor \geq 2$ by assumption, and $1 - \frac{\lfloor k/\ell \rfloor}{n} \geq 1 - 1/\ell \geq 1/2$.
\endproof
\begin{proposition}\label{prop:il(ell)-fairness}
For any \(\ell \in \N\) and any instance such that \(\max_{G\in\G} |G| \le \ell \), every dominant strategy equilibrium outcome of the Individual Lottery with limit \(\ell\) is \(1/\ell-\)fair.
\end{proposition}
\proof[Proof of Proposition~\ref{prop:il(ell)-fairness}]
We construct a random order over agents $\rOrder$ using Algorithm~\ref{alg:simultaneous}: set $S = G_{i} \cup G_{j}$, and independently generate (i) a uniform random order \(\rOrder^S\) over $S$, (ii) a uniform random order \(\rOrder^-\) over \(\sN \setminus S\), and (iii) uniform random positions $\positions \subseteq \{1, \ldots, |\sN|\}$ where agents in $S$ will be placed. By Lemma~\ref{lem:uniform-random-order}, the resulting order $\rOrder$ is uniformly distributed.

Without loss of generality, we assume
\begin{equation} \ell \geq |G_i| \geq |G_j|.\label{eq:size-bounds}\end{equation}
We let
\begin{align}
\tau_i(\rOrder^{-}) & = \tau(k - |G_{i}| + 1, \rOrder^-),\\
\tau_j(\rOrder^{-}) & = \tau(k - |G_{j}| + 1, \rOrder^-) - \tau(k - |G_{i}| + 1, \rOrder^-),
\end{align}
be as defined in \eqref{eq:def:tau} where the size of each agent is its request, that is, \(|\order_t| = a_{\order_t}\). Note that by definition,
\begin{equation}
1\leq \tau_i(\rOrder^{-}),\hspace{.2 in}   \text{ and } \hspace{.2 in} \tau_j(\rOrder^-) \leq |G_i|-|G_j|. \label{eq:taubounds}
\end{equation}
In addition, for $s \in \{1, \ldots, |S|\}$ let $T_s(\positions)$ be the $s^{th}$ smallest value in $\positions$, so $T_1(\positions)$ denotes the first position of $\rOrder$ containing a member of $G_i \cup G_j$. Note that
\[\bP(T_1(\positions) =  t) =  \left(\frac{|G_i|+|G_{j}|}{n}\right) \frac{ \binom{n-t}{|G_i| +|G_j|-1}}{ \binom{n-1}{|G_i| +|G_j|-1}}, \]
which is decreasing in $t$. From this, it follows that for any $\rOrder^-$,
\begin{equation} \frac{\bP(T_1 \leq \tau_i + \tau_j \vert \rOrder^-)}{\bP(T_1 \leq \tau_i \vert \rOrder^-)}\leq \frac{\tau_i+\tau_j}{\tau_i} \leq 1 +|G_i|-|G_j|,  \label{eq:tauratio}\end{equation}
where the second inequality comes from \eqref{eq:taubounds}. Our final definition is to let
\begin{equation} A_i  = \{ \rOrder^S_1 \in G_i\}, \hspace{.5 in} A_j  = \{ \rOrder_1^S \in G_j\}, \end{equation}
and note that
\begin{equation} \mathbb{P}(A_i) = \frac{|G_i|}{|G_i| + |G_j|} = 1 - \mathbb{P}(A_j).\label{eq:aiprob} \end{equation}

Definitions out of the way, we proceed with the proof. Note that
\[ \frac{u_j(\IL(\a))}{u_i(\IL(\a))} = \frac{\E[\E[u_j(x(\a,\rOrder))\vert \rOrder^-]]}{\E[\E[u_i(x(\a,\rOrder))\vert \rOrder^-]]} \leq \max_{\order^-} \frac{\E[u_j(x(\a,\rOrder))\vert \rOrder^-=\order^-]}{\E[u_i(x(\a,\rOrder))\vert \rOrder^-=\order^-]}.\]
Therefore, to establish $1/\ell$ fairness, it suffices to show that for every $\rOrder^-$,
\begin{equation} \frac{1}{\ell} \leq  \frac{\E[u_j(x(\a,\rOrder))\vert \rOrder^-]}{\E[u_i(x(\a,\rOrder))\vert \rOrder^-]} \leq \ell. \label{eq:conditional-bounds}\end{equation}
We claim that
\begin{equation}
\bP(A_i) \bP(T_1 \leq \tau_i \vert \rOrder^-) \leq \E[u_i(x(\a,\rOrder)) \vert \rOrder^-] \leq \bP(T_1 \leq \tau_i \vert \rOrder^-). \label{eq:ui-bounds}
\end{equation}
The left inequality follows because whenever $T_1 \leq \tau_i$ and a member of $G_i$ comes before all members of $G_j$, group $G_i$ gets a payoff of $1$. The right inequality follows because the definition of $\tau_i$ ensures that $G_i$ can get a payoff of one only if a member of $G_i$ is in the first $\tau_i$ positions of $\rOrder$. By analogous reasoning, we have
\begin{equation}
\bP(A_j) \bP(T_1 \leq \tau_i \vert \rOrder^-) \leq \E[u_j(x(\a,\rOrder)) \vert \rOrder^-] \leq \bP(T_1 \leq \tau_i+\tau_j \vert \rOrder^-) - \bP(A_i)\bP(\tau_i < T_1 \leq \tau_i+\tau_j \vert \rOrder^-), \label{eq:uj-bounds}
\end{equation}
where the right inequality follows because in order for group $G_j$ to get utility one, we must have $T_1 \leq \tau_i + \tau_j$, and if $T_1 \in (\tau_i, \tau_i+\tau_j]$, then a member of $G_j$ must appear before all members of $G_i$.

We now prove the upper-bound in \eqref{eq:conditional-bounds}. Combining \eqref{eq:ui-bounds} and \eqref{eq:uj-bounds}, we see that
\begin{align}
\frac{\E[u_j(x(\a,\rOrder)) \vert \rOrder^-]}{\E[u_i(x(\a,\rOrder)) \vert \rOrder^-]} & \leq \frac{\bP(T_1 \leq \tau_i+\tau_j \vert \rOrder^-) - \bP(A_i)\bP(\tau_i < T_1 \leq \tau_i+\tau_j \vert \rOrder^-)}{\bP(A_i) \bP(T_1 \leq \tau_i \vert \rOrder^-)}  \nonumber\\
& =  \frac{\bP(T_1 \leq \tau_i+\tau_j\vert \rOrder^-)(1 - \bP(A_i)) + \bP(A_i) \bP( T_1 \leq \tau_i\vert \rOrder^-)}{\bP(A_i) \bP(T_1 \leq \tau_i\vert \rOrder^-)}  \nonumber \\
& = \frac{|G_j|}{|G_i|}\frac{\bP(T_1 \leq \tau_i+\tau_j\vert \rOrder^-)}{ \bP(T_1 \leq \tau_i\vert \rOrder^-)} + 1  \nonumber \\
& \leq \frac{|G_j| - |G_j|^2}{|G_i|} + |G_j| + 1 \nonumber \\
& \leq \ell. \label{eq:ujui}
\end{align}
The second inequality uses \eqref{eq:tauratio}. The final inequality follows because if $|G_j| =  \ell$, then $|G_i| = \ell$ by \eqref{eq:size-bounds}, and thus the expression is equal to $2$;\footnote{We assume $\ell \geq 2$ because if $\ell = 1$ and all groups have size one, the individual lottery simply selects $k$ agents uniformly at random, and is perfectly fair.} if $|G_j| < \ell$, then the expression is at most $\ell$ because $\frac{|G_j| - |G_j|^2}{|G_i|} \leq 0$.

Meanwhile,~\eqref{eq:ui-bounds} and~\eqref{eq:uj-bounds} also imply that
\begin{equation}
\frac{\E[u_j(x(\a,\rOrder)) \vert \rOrder^-]}{\E[u_i(x(\a,\rOrder)) \vert \rOrder^-]} \geq \bP(A_j) = \frac{|G_j|}{|G_i|+|G_j|}. \label{eq:uiuj}
\end{equation}
If $|G_i| < \ell$ or $|G_j| > 1$, the ratio on the right is at least $1/\ell$, and the proof is complete. Thus, all that remains is to show that the lower bound in \eqref{eq:conditional-bounds} holds when $|G_i| = \ell$ and $|G_j| = 1$.

Our analysis will condition on both $\rOrder^-$ and $\positions$. We note that
\[\E[u_i(x(\a,\rOrder)) \vert \rOrder^-, \positions] > 0 \Leftrightarrow T_1(\positions) \leq \tau_i(\rOrder^-).\]
Therefore,
\[ \frac{\E[u_j(x(\a,\rOrder))\vert \rOrder^-]}{\E[u_i(x(\a,\rOrder))\vert \rOrder^-]} =  \frac{\E_\positions[\E[u_j(x(\a,\rOrder))\vert \rOrder^-,\positions]]}{\E_\positions[\E[u_i(x(\a,\rOrder))\vert \rOrder^-,\positions]]}\geq \min_{\positions : \,\, T_1(\positions) \leq \tau_i(\rOrder^-)} \frac{\E[u_j(x(\a,\rOrder))\vert \rOrder^-,\positions]}{\E[u_i(x(\a,\rOrder))\vert \rOrder^-,\positions]}.\]
We will show that the quantity on the right is at least $1/\ell$. To do this, we let
\begin{equation} \tau_{ij}(\rOrder^-) = \tau(k-|G_{j}|-|G_i| + 1,\rOrder^-),\end{equation}
be as defined in~\eqref{eq:def:tau} with size function \(|\order_t| = a_{\order_t}\). If $T_2(\positions) \leq \tau_{ij}(\rOrder^-)$, then because each agent requests at most $\ell = |G_i|$ tickets, agent $j$ will receive utility of one if first or second in $\rOrder^S$. Thus,
\[\frac{\E[ u_j(x(\a,\rOrder))\vert \rOrder^-, \positions]}{\E[ u_i(x(\a,\rOrder))\vert \rOrder^-, \positions]} \geq \E[ u_j(x(\a,\rOrder))\vert \rOrder^-, \positions] \geq \frac{2}{\ell+1} \geq \frac{1}{\ell}.\]
Meanwhile, if $T_2(P) > \tau_{ij}(\rOrder^-)$, then each group gets utility $1$ only if one of its members is first in $\rOrder^S$. In this case,
\[\frac{\E[ u_j(x(\a,\rOrder))\vert \rOrder^-, \positions]}{\E[ u_i(x(\a,\rOrder))\vert \rOrder^-, \positions]}= \frac{\bP(A_j)}{\bP(A_i)} = \frac{1}{\ell}.\]
\endproof

\newpage
\section{\NameProposedMechanism}\label{app:spl}
\subsection{Incentives}
\begin{proposition}\label{prop:spl-random-order}
Algorithm~\ref{alg:exponentials} generates a random order $\rOrder \in \o_{\sN}$  distributed according to~\eqref{eq:size_proportional_lottery} conditional on $\a$.
\end{proposition}
\proof[Proof of Proposition~\ref{prop:spl-random-order}]
Fix any order \(\order\) over \(\sN\). Let $Y_j = a_j X_j$. It follows that $\bP(Y_j> t) = e^{-t/a_j}$, so each $Y_j$ is distributed as an exponential random variable with mean $a_i$. Moreover, the $Y_j$ are independent. Let $\rOrder$ be the order generated by the algorithm. We have that
\[\mathbb{P}({\Sigma_1} = j) = \mathbb{P}(Y_j = \min_{i \in \sN} Y_i) = \frac{1/a_j}{ \sum_{i \in \sN}1/a_i},\]
where the second equality follows from well-known properties of the minimum of exponential random variables.\footnote{See e.g. \url{https://en.wikipedia.org/wiki/Exponential_distribution}.} Furthermore, the definition of $\Sigma$ and the memoryless property of exponential random variables imply that for $t \in \{1, \ldots, n\}$ and $j \not \in \Sigma_{[t-1]}$,
\[\mathbb{P}({\Sigma_{t}} = j \vert \Sigma_{[t-1]}) = \mathbb{P}(Y_j = \min_{i \in \sN \backslash \Sigma_{[t-1]}} Y_i) = \frac{1/a_j}{ \sum_{i \in \sN \backslash \Sigma_{[t-1]}}1/a_i}.\]
This implies that
\[\Pr(\Sigma = \sigma) = \prod_{t = 1}^{n} \mathbb{P}({\Sigma_{t}} = \sigma_{t} \vert \Sigma_{[t-1]} = \sigma_{[t-1]}) = \prod_{t = 1}^{n} \frac{1/a_{\sigma_t}}{ \sum \limits_{i \in \sN \backslash \sigma_{[t-1]}}1/a_i},\]
as claimed.
\endproof
\proof[Proof of Proposition~\ref{prop:SPL-incentives}]
We start by proving that agents have no incentives to request more tickets than their group size. Formally, if we let $i \in G$, $a_i = |G|$ and $a'_i > |G|$ then for every action profile \(\a_{-i} \in A_{-i}\),
\[u_i(\SPL(a_i,\a_{-i})) \ge u_i(\SPL(a'_i,\a_{-i})).\]
This follows because the set of orders over agents in which \(G\) get a payoff of \(1\) is the same under both strategies, and by reducing its request agent \(i\) improves her probability of being drawn early.

We now show that if group \(G\) is such that \(|G| \le 3\), then selecting group request $\a_G$ is dominant for \(G\). Given an action profile $\a \in A$, we generate a random order over agents $\rOrder$ using the Algorithm~\ref{alg:exponentials}: we draw iid exponential random variables $X_i$ for each agent $i$, and sort agents in increasing order according to $a_iX_i$. From Proposition~\ref{prop:spl-random-order}, it follows that \(\sq\) is distributed according to~\eqref{eq:size_proportional_lottery} conditional on $\a$. Let \(T\) be as in Definition~\eqref{eq:T-def}, intuitively \(T\) is the score threshold that some members of \(G\) must clear in order to ensure the group a payoff of \(1\).
Furthermore, when \(G\) is selecting the group request strategy, it will get a payoff of \(1\) if and only if at least one of its members has a score lower than \(T\),  that is, $\min_{i\in G}\{a_i X_i\} < T$. Because $\min_{i\in G}\{a_i X_i\} \sim Exp(1)$, it follows that for $i\in G$,
\begin{equation}\label{eq:spl-gr-cond-utility}
\E[u_i(\SPL(\a_G,\a_{-G})) | T] = \bP(\min_{i\in G}\{a_i X_i\} < T) = 1 -e^{-T}.
\end{equation}
% \cb{This equation might be moved to the body.}
Because \(T\) is independent of the strategy followed by \(G\), it suffices to show that for any deviation $\a'_G$ the conditional expected utility of $G$ given $T$ is less than or equal the right side of~\eqref{eq:spl-gr-cond-utility}.

We have already established that it is never beneficial for agents to request more tickets than their group size. Hence, without loss of generality we assume that each member of \(G\) will request at most \(|G|\) tickets.

If $|G|=1$, then the group request is the only feasible strategy so it is dominant.

If $|G|=2$, then the only deviations we need to consider are $\a'_G = (1,2), (1,1)$. The first strategy is dominated by the group request, because the allocation of the member requesting \(1\) ticket is irrelevant for the outcome of group \(G\). Under the second strategy, $G$ gets a payoff of \(1\) if and only if both members have a score lower than \(T\). In particular, agent \(i \in G\) must have a score lower than \(T\). This happens with probability
\[\bP(a_i' X_i < T) = \bP(X_i < T) =  1 -e^{-T}.\]
Note that the quantity above coincides with~\eqref{eq:spl-gr-cond-utility},  implying that the utility of \(G\) when selecting $\a'_G = (1,1)$ is at most its utility under the group request strategy.

If $|G| = 3$, there are \(27\) feasible strategies (26 deviations from the group request), but by symmetry we only need to evaluate 9 of them:
\[\a'_G = (1,1,1), (1,1,2), (1,1,3), (1,2,2), (1,2,3), (1,3,3), (2,2,2), (2,2,3), (2,3,3).\]
We argue now that the group request dominates all strategies above in which there is at least one agent requesting \(1\) ticket. Note that under any of these strategies, \(G\) will get a payoff of \(1\) only if the remaining \(2\) members are awarded two or more tickets. From the case \(|G| = 2\), we know that the probability of this event is at most the right hand side of~\eqref{eq:spl-gr-cond-utility}. This implies that the group request strategy dominates all these deviations.

There are only 3 strategies remaining: \(\a'_G = (3,3,2), (3,2,2), (2,2,2)\). The first strategy is dominated by the group request, because the allocation of the member requesting \(2\) tickets is irrelevant for the outcome of group \(G\). The second strategy is dominated by \((3,2,1)\). This follows because the set of orders over agents in which \(G\) get a payoff of \(1\) is the same under both strategies, and by reducing its request the last agent improves her probability of being drawn early. A similar argument shows that the last strategy is dominated by \((1,2,2)\).
\endproof
\proof[Analysis of Example \ref{ex:SPL-incentives}]\label{ex:SPL-incentives-proof}
Let \(i\) be a member of the large group. We let \(\a = (\a_{G_i}, \a_{-{G_i}})\) denote the group request action profile, and \(\a'_{G_i}\) denote the strategy where all members of \(G_i\) request \(2\) tickets. We will show that for \(n\ge 17\),
\[u_i(\SPL(\a'_{G_i},\a_{-{G_i}})) \ge u_i(\SPL(\a_{G_i},\a_{-{G_i}})).\]
Let \(m=n-3\) be the number of groups. We claim that
\begin{align}
u_i(\SPL(\a_{G_i},\a_{-{G_i}})) &= 1 - \frac{1}{m}\label{eq:spl-ex-gr},\\
u_i(\SPL(\a'_{G_i},\a_{-{G_i}})) &= 1 - \sum_{t=1}^m\left(\prod_{i=1}^{t-1}\frac{m - i}{m + 2 - i}\right)\left(\frac{2}{m + 2 - t}\right)\left(\prod_{i=t}^{m-1}\frac{m - i}{m + 3/2 - i}\right).\label{eq:spl-ex-de}
\end{align}
This implies our result as for \(m\ge 14\) the expression in~\eqref{eq:spl-ex-de} exceeds the expression in~\eqref{eq:spl-ex-gr}.

First, we will show~\eqref{eq:spl-ex-gr}. Because \(G_i\) is selecting the group request strategy, it will get a payoff of \(0\) if and only if all agents from small groups are processed before its members. This event happens with probability
\[\prod_{i=1}^{m-1}\frac{m - i}{m + 1 - i} = \frac{1}{m}.\]
Secondly, we show~\eqref{eq:spl-ex-de}. If all members of \(G_i\) are requesting \(2\) tickets, then \(G_i\) will get a payoff of \(0\) if and only if three of its members are processed after all agents in small groups. Moreover, the probability that at step \(t=1,\ldots, m\) a member of \(G_i\) is processed for the first time is
\begin{equation}\label{eq:spl-ex-t1}
\left(\prod_{i=1}^{t-1}\frac{m - i}{m + 2 - i}\right)\left(\frac{2}{m + 2 - t}\right).
\end{equation}
In the expression above we used that \(\sum_{i\in G} 1/a_i' = 2\), and that at the beginning of step \(i \le t\), there are \(m + 2 - i\) agents in small groups that have not been processed yet. Note that if \(t=1\), then the expression above reduces to the probability of processing a member of the large group at the first step, that is, \(2/(m + 1)\).

Finally, the probability that all the remaining agents in small groups are processed before the three remaining members of the large group is
\begin{equation}\label{eq:spl-ex-t2}
\prod_{i=t}^{m-1}\frac{m - i}{m + 3/2 - i}.
\end{equation}
Here we are using that if \(j\in G\) was processed at step \(t\), then \(\sum_{i\in G\setminus \{j\}} 1/a_i' = 3/2\). Note that if \(t=m\), then the expression above is \(1\).

Multiplying~\eqref{eq:spl-ex-t1} by~\eqref{eq:spl-ex-t2} and summing all possible values of \(t\) yields~\eqref{eq:spl-ex-de}.
\endproof
\subsubsection{Proof of Proposition~\ref{prop:spl-gr-optimality-restricted}}
\begin{lemma}\label{lemma:threshold}
Group \(G\) gets a payoff of \(1\) if and only if
\begin{equation} \sum_{i \in G} a_i {\bf 1}( a_i X_i < T) \geq |G|. \label{eq:success} \end{equation}
\end{lemma}
\proof[Proof of Lemma~\ref{lemma:threshold}]
First, suppose that~\eqref{eq:success} holds. From the definition of \(T\) in~\eqref{eq:T-def}, it follows that at most \(k-|G|\) tickets are allocated to agents not in \(G\) who have a score lower than \(T\). Furthermore, as~\eqref{eq:success} holds it must be the case that the sum of the requests of agents in \(G\) who have a score lower than \(T\) is at least \(|G|\). Therefore, group \(G\) is awarded \(|G|\) or more tickets.

Conversely, suppose that~\eqref{eq:success} does not hold. We will consider two cases:
\begin{itemize}
    \item[(i)]Only agents with score lower than \(T\) are awarded.
    \item[(ii)]There are agents with score \(T\) or higher that are awarded.
\end{itemize}
Assume first that (i) holds. Then as~\eqref{eq:success} doesn't hold individuals in \(G\) must receive fewer than \(|G|\) tickets.

Assume now that (ii) holds. From the definition of \(T\) in~\eqref{eq:T-def}, it follows that individuals not in \(G\) must receive strictly more than \(k - |G|\) tickets. This implies that individuals in \(G\) must receive fewer than \(|G|\) tickets.
\endproof
\begin{lemma}\label{lemma:spl-reciprocals-ub}
Fix an arbitrary group \(G\). Let \(r \in \{0,\ldots,|G|-1\}\). For every strategy \(\a_G\in\B_r\), it follows that
\begin{align}
\sum_{i \in G}\frac{1}{a_i} \le r + 1.\label{eq:lambda-ub}
\end{align}
\end{lemma}
\proof[Proof of Lemma~\ref{lemma:spl-reciprocals-ub}] For simplicity, we shall assume that \(G = \{1,\ldots,s\}\) and \(a_1 \le a_2 \le \cdots \le a_s\). Thus, a strategy \(\a_G \in \N^s\) is in \(\B_r\) if and only if
\begin{align}\label{eq:br-characterization}
\sum_{i=1}^{r+1} a_{i} &\ge s \text{ and }
\sum_{i=s-r+1}^s a_{i} \le s - 1.
\end{align}
Consider the following optimization problem:
\begin{equation}\label{reciprocals-max}
\begin{array}{lll}
\max &\sum_{i=1}^s1/a_i\\
\subjectto &\sum_{i=1}^{r+1} a_{i} \ge s\\
&\sum_{i=s-r+1}^s a_{i} \le s - 1.\\
&1\le a_1 \le \cdots \le a_{s}
\end{array}
\end{equation}
Note that from~\eqref{eq:br-characterization} it follows that every strategy \(\a_G \in \B_r\) is a feasible solution for this problem. Therefore, to prove~\eqref{eq:lambda-ub} it suffices to show that the optimal value of this problem is at most \(r+1\).

We start by proving that an optimal solution \(\a_G^*\) of~\eqref{reciprocals-max} must satisfy
\begin{align}\label{reciprocals-max-cond}
a_j^* &= s - \sum_{i=1}^r a_i^* \text{ for every }j= r+1,\ldots,s.
\end{align}
Suppose \(\a\in \B_r\) is such that \(a_j > s + \sum_{i=1}^r a_i \) for some \(j\in\{r+1,\ldots,s\}\). If we replace \(a_j\) by \(a_j' = s - \sum_{i=1}^r a_i\) then we increase the objective value as \(1/a_j < 1/a_j'\). Moreover, \(\a'\) will still be in \(\B_r\) as \(\sum_{i=1}^{r+1} a_{i}' = \sum_{i=1}^r a_i + a_j' = s\) and \(\sum_{i=s-r+1}^s a_i' \le \sum_{i=s-r+1}^s a_i \le s - 1\). The last inequality follows as \(\a\in \B_r\).

It follows from~\eqref{reciprocals-max-cond} that we can incorporate in~\eqref{reciprocals-max} the constraints \[a_{r+1} = \cdots = a_s = s - \sum_{i=1}^r a_i,\] without decreasing the optimal value. Moreover, if we remove the constraint \(\sum_{i=s-r+1}^s a_{i} \le s - 1\) then the optimal value will be higher or the same. By including both modifications we obtain the following relaxation of~\eqref{reciprocals-max}:
\begin{equation}\label{reciprocals-relax}
\begin{array}{lll}
\max &\sum_{i=1}^r\frac{1}{a_i} + \frac{(s-r)}{a_{r+1}}\\
\subjectto &\sum_{i=1}^{r+1} a_i = s\\
&1\le a_1 \le \cdots \le a_{r+1}
\end{array}
\end{equation}
Clearly an optimum of~\eqref{reciprocals-relax} exists as the objective function is continuous and the feasible set is non-empty and compact. Moreover, we are maximizing a convex function on a convex set then there exists a globally optimal solution that is an extreme point of the feasible set. The extreme points of the feasible set are \(\a^0,\ldots, \a^{r}\), where
\begin{align*}
1 = a^j_1 = \cdots = a^j_j,\quad \frac{s - j}{r+1-j} = a^j_{j+1} = \cdots = a^j_{r+1}.
\end{align*}
Furthermore, the objective value evaluated at any extreme point is equal to \(r+1\). To see this note that objective value at \(\a^j\) is
\[j + (s - j) \left(\frac{r+1-j}{s-j}\right)= r + 1.\]
Therefore, the optimal value of~\eqref{reciprocals-relax} is \(r+1\). Because~\eqref{reciprocals-relax} is a relaxation of~\eqref{reciprocals-max}, it follows that the optimal value of~\eqref{reciprocals-max} is at most \(r+1\).
\endproof
\proof[Proof of Proposition~\ref{prop:spl-gr-optimality-restricted}]
Let \( r \in \{0,\ldots, s-1\} \). We formulate the problem of finding the strategy in \(\B_r\) that maximizes the expected payoff of \(G\) given the threshold \(T\) as a programming problem. From Lemma~\ref{lemma:threshold} and since we are considering only strategies in \(\B_r\), it follows that group \(G\) will get a payoff of \(1\) if and only if there are \(r + 1\) or more agents with a score lower than \(T\). For each agent \(i \in G\), we let \(B_i\) be a random variable that indicates if the score of agent \(i\) is lower than \(T\), more precisely, \(B_i = \mathbf{1}(a_iX_i < T)\). Observe that given \(T\) and any action \(a_i\), because \(X_i \sim Exp(1)\) then \(B_i \sim Bernoulli (1 - e^{-T/a_i})\). Hence, our formulation is
\begin{equation}\label{agent-based-formulation}
\begin{array}{lll}
\max & \bP(\sum_{i\in G}  B_i\ge r + 1)\\
\subjectto &B_i \sim Bernoulli(1-e^{-T/a_i})&\forall i \in G\\
&\a_G\in\B_r
\end{array}
\end{equation}
Let \(Z_i\) be the Poisson random variable of rate \(T/a_i\). Note that \(Z_i\) first-order stochastically dominates \(B_i\). Hence, the following problem is a relaxation of~\eqref{agent-based-formulation}.
\begin{equation}\label{eq:poisson-problem}
\begin{array}{lll}
\max & \bP(\sum_{i\in G}  Z_i\ge r + 1)\\
\subjectto &Z_i \sim Poisson(T/a_i)&\forall i \in G\\
&\a_G\in\B_r
\end{array}
\end{equation}
Using that the sum of independent Poisson random variables is Poisson-distributed, we have that \(\sum_{i\in G} Z_i \sim Poisson(\sum_i T/a_i)\). Moreover, if \(X\sim Poisson(\lambda)\) then~\citet{johnson2005univariate} state the following bound:
\begin{equation}
\bP(X\ge x) \le~1 - e^{-\lambda/x},\quad x \ge \lambda.\label{poisson-tail-bound}
\end{equation}
If \(T\le 1\), then Lemma~\ref{lemma:spl-reciprocals-ub} implies
\begin{equation}\sum_{i\in G} \frac{T}{a_i}  \le \sum_{i \in G} \frac{1}{a_i} \le r+1.\end{equation}
Therefore, we can apply~\eqref{poisson-tail-bound} to obtain
\begin{equation}\bP(\sum_{i\in G} Z_i  \ge r + 1) \le 1 - e^{-(\sum_i T/a_i)/(r+1)} \le 1 - e^{-T}.\end{equation}
The last inequality follows from Lemma~\ref{lemma:spl-reciprocals-ub}.

From~\eqref{eq:spl-GT-problem-gr} we have that \(1 - e^{-T}\) correspond to the utility of \(G\) under the group request strategy. This implies our result as the optimal value of the relaxation~\eqref{eq:poisson-problem} is at most the utility under the group request strategy. %All that remains is to prove our claim. Under the group request strategy \(G\) gets a payoff of \(1\) if and only if at least one of its members has a score lower than \(T\). Furthermore, under this strategy \(\min_{i\in G} a_iX_i\) is distributed as \(Exp(1)\), so the probability of success under a group request strategy is \(1 - e^{-T}\).
\endproof
\subsection{Performance}
\subsubsection{Proof of Theorem~\ref{thm:spl-performance}}
\proof[Proof of Theorem~\ref{thm:spl-performance}]
In this proof, whenever we study a mechanism we assume that the action profile selected \(\a\) is its corresponding group request strategy.

We start by proving the efficiency guarantee. From Proposition~\ref{prop:mech-dominance}, we have that for any instance the utilization under the {\NameProposedMechanism} is at least the utilization under the Group Request with Replacement. This can be seen as
\begin{equation}
U(\SPL(\a)) = \frac{\sum_{i \in \sN}u_i(\SPL(\a))}{k} \ge \frac{\sum_{i \in \sN}u_i(\GLR(\a))}{k} = U(\GLR(\a)).
\end{equation}
The inequality follows from~\eqref{eq:mech-dominance}. Therefore, it suffices to show that for any instance in \(I(\kappa,\alpha)\), the Group Lottery with Replacement is \((1 - \kappa)g(\alpha)\)-efficient. This follows immediately by Lemma~\ref{lemma:glr-utility-lb-body}:
\begin{equation}
 U(\GLR(\a)) = \frac{\sum_{i \in \sN}u_i(\GLR(\a))}{k} \ge \frac{n\left(\frac{k}{n}(1 - \kappa) g(\alpha)\right)}{k} = (1-\kappa)g(\alpha).
\end{equation}
Now we turn to the fairness guarantee. From Proposition~\ref{prop:mech-dominance} , we have that for any instance and any pair of agents \(i,j\),
\begin{equation}
\frac{u_i(\SPL(\a))}{u_j(\SPL(\a))} \ge \frac{u_i(\GLR(\a))}{u_j(\GL(\a))}.
\end{equation}
Moreover, combining Lemma~\ref{lemma:glr-utility-lb-body} and Lemma~\ref{lemma:gl-utility-bounds} yields
\begin{equation}
\frac{u_i(\GLR(\a))}{u_i(\GL(\a))} \ge \frac{\frac{k}{n}(1 - \kappa) g(\alpha)}{\frac{k}{n} \left(1 + \kappa\right)} \ge (1-\kappa)^2g(\alpha) \ge (1-2\kappa)g(\alpha).
\end{equation}
The second inequality follows from the fact that for any \(x \ge 0\),
\[\frac{1}{1+x} \ge 1-x.\]
\endproof
\subsubsection{Proof of Lemma~\ref{prop:mech-dominance}}\label{app:mech-dominance}
Let $\s_{\sN}$ be the set of finite sequences of agents and draw the random sequence $\sq \in \mathcal{S}_{\sN}$ by letting $\sq_t$ be iid with
\begin{equation}\label{eq:agent-seq-dist}
\bP (\sq_t=i) = \frac{1/|G_i|}{\sum_{j \in \sN}1/|G_j|},
\end{equation}
stopping once all agents have been drawn at least once, that is, for each \(i \in \sN\) there exists \(t\) such that \(i = \sq_{t}\). This occurs with probability one, implying that this procedure generates a valid distribution over $\mathcal{S}_{\sN}$.

Define $\order^{GR} : \s_{\sN} \to  \s_{\G}$ by
\begin{equation} \order_j^{GR}(\Sigma) = G_{\sq_j}.\label{eq:sigma-tilde} \end{equation}

Define \(\order^{IW}: \mathcal{S}_{\sN} \to \o_{\sN}\) by
 \begin{align}
T^{IW}_j(\sq) & = \min\{t \in \N: | \sq_{[t]}| = j\},\nonumber \\
\order^{IW}_j(\sq) & = \sq_{T^{IW}_j(\Sigma)}. \label{eq:sigma-iw}
\end{align}
Note that for each $\Sigma \in \s_{\sN}$ and each $t \in \N$, $\order_{[t]}^{GR}(\Sigma)\subseteq \G$. Define \(\order^{GL}: \mathcal{S}_{\sN} \to \o_\G\) by
\begin{align}
T^{GL}_j(\sq) &  = \min\{t \in \N: |\order_{[t]}^{GR}(\Sigma)| = j\}. \nonumber \\
\order^{GL}_j(\sq) & = \order_{T^{GL}_j(\Sigma)}^{GR}(\Sigma). \label{eq:sigma-gl}
\end{align}

\begin{proposition}\label{prop:coupling-dist}
Let \(\order^{GR}, \order^{IW}, \order^{GL}\) be as in \eqref{eq:sigma-tilde}, \eqref{eq:sigma-iw}, \eqref{eq:sigma-gl}. If $\sq$ is drawn according to \eqref{eq:agent-seq-dist}, then
\begin{itemize}
    \item \(\order^{GR}(\sq)\) a sequence of \(k\) elements in \(\G\), where each element is independently and uniformly sampled with replacement from \(\G\).
    \item \(\order^{IW}(\sq)\) is an order over \(\sN\) distributed as in~\eqref{eq:size_proportional_lottery} given a group request action profile.
    \item \(\order^{GL}(\sq)\) is a uniform order over \(\G\).
\end{itemize}
\end{proposition}
\proof[Proof of Proposition~\ref{prop:coupling-dist}]
From the definition of \(\order^{IW}(\sq)\), we know that it skip every agent in \(\sq\) that has already appeared. Hence, we are sequentially sampling agents without replacement, with probability inversely proportional to the size of its groups. Therefore, it correspond to an order over agents distributed according to~\eqref{eq:size_proportional_lottery} when each agent is requesting its group size.

From~\eqref{eq:agent-seq-dist} it follows that for each $G \in \G$ and each $t$, $\bP (\sq_t\in G) = 1/|\G|$. That is, the marginal distribution over groups is uniform. It immediately follows from the definition of \(\order^{GLR}(\sq)\) that it is sampling groups uniformly at random with replacement. Moreover, from the definition of \(\order^{GL}(\sq)\) we know that it skip every agent in \(\sq\) whose group has already appeared. Therefore, we are sampling groups uniformly at random without replacement, generating a uniform order over \(\G\).
\endproof
\proof[Proof of Lemma~\ref{prop:mech-dominance}]
Let $\sq$ be drawn according to \eqref{eq:agent-seq-dist}, and \(\order^{GR}, \order^{IW}, \order^{GL}\) be as in \eqref{eq:sigma-tilde}, \eqref{eq:sigma-iw}, \eqref{eq:sigma-gl}. From Proposition~\ref{prop:coupling-dist} it follows that
\begin{align*}
u_i(\GLR(\a)) &=  \E[u_i(\xgl(\a,\order^{GR}(\sq)))],\\
u_i(\SPL(\a)) &=  \E[u_i(\xspl(\a,\order^{IW}(\sq)))],\\
u_i(\GL(\a)) &=  \E[u_i(\xgl(\a,\order^{GL}(\sq)))].
\end{align*}
Therefore, it suffices to show that for any realization of $\sq$,
\begin{align}\label{eq:utilities-dominance}
u_i(\xgl(\a,\order^{GR}(\sq))) \le
u_i(\xspl(\a,\order^{IW}(\sq))) \le
u_i(\xgl(\a,\order^{GL}(\sq))).
\end{align}
Observe that given \(\sq\), each of the utilities above is either \(0\) or \(1\). Hence, to prove~\eqref{eq:utilities-dominance} we will show that: (i) if the utility of \(i\) under the Group Lottery with Replacement is 1 then its utility under the {\NameProposedMechanism} is also 1, and (ii) if the utility of agent \(i\) under the {\NameProposedMechanism}  is 1 then its utility under the Group Lottery is also 1. Because agents are playing the group request strategy, whenever a group or agent is being processed, it is given a number of tickets equal to the minimum of its group size and the number of remaining tickets.

If the utility of \(i\) under the Group Lottery with Replacement is \(1\), then the number of tickets allocated before \(G_{i}\) is processed is at most \(k - |G_i|\). Formally, if we let $t$ be the first time at which a member of $G_i$ appears in $\sq$, then
\begin{equation}\label{eq:glr-allocated-objs}
\sum_{j=1}^{t-1} |G_{\sq_j}| \le k - |G_i|.
\end{equation}
In the left hand side, we use that the sequence of groups \(\order^{GR}(\sq)\) is determined by replacing each agent in \(\sq\) by its group. In contrast, in the {\NameProposedMechanism}, the order over agents \(\order^{IW}(\sq)\) is constructed by skipping all agents in \(\sq\) that have already appeared. Hence, in this mechanism the number of tickets allocated before $\Sigma_t$ appears in $\order^{IW}(\Sigma)$ is the same or lower than the left hand side of~\eqref{eq:glr-allocated-objs}. This implies that the utility of \(i\) under the {\NameProposedMechanism} is also \(1\).

Meanwhile, suppose that the utility of \(i\) under the {\NameProposedMechanism} is \(1\), then the number of tickets allocated before $\Sigma_t$ appears in $\order^{IW}(\Sigma)$ is at most \(k - |G_i|\). In the Group Lottery, the order over groups \(\order^{GL}(\sq)\) is constructed by replacing each agent in \(\sq\) by its group, and skipping all groups that have already appeared. Note that each time \(\order^{IW}(\sq)\) skips an agent \(\sq_j\), \(\order^{GL}(\sq)\) also skips the group \(G_{\sq_j}\). Therefore, the number of tickets allocated before $G_{\Sigma_t}$ appears in $\order^{GL}(\Sigma)$ is the same or lower than the number of tickets allocated before $\Sigma_t$ appears in $\order^{IW}(\Sigma)$. Implying that the utility of agent \(i\) under the Group Lottery is also 1.
\endproof
\subsubsection{Proof of Lemma~\ref{lemma:glr-utility-lb-body}}\label{app:glr}
\begin{proposition}\label{prop:glr-random-order}
Given a set $V$ of \(m\) elements and a natural number \(k\), the following algorithm generates a sequence \(\rOrder\) of \(k\) elements where \(\rOrder_t\) is independent and uniformly sampled from \(V\):
\begin{enumerate}
\item Select an element \(G\) of \(V\).
\item Generate a sequence of \(k\) elements \(\rOrder^{-}\), where \(\rOrder_t^{-}\) is independently and uniformly draw from \(V\setminus G\).
\item Generate a sequence of \(k\) independent binary random variables \(X\), where  \(X_t\sim Bernoulli(1/m)\).
\item For \(t=1,\ldots,k\) set
\[\rOrder_t =\begin{cases} G &\text{if }X_t = 1,\\ \rOrder^{-}_t &\text{otherwise}.\end{cases}\]
\end{enumerate}
\end{proposition}
\proof[Proof of Proposition~\ref{prop:glr-random-order}]
Observe that \(\rOrder_t\) depends only on \(X_t\) and \(\rOrder^{-}_t\), hence, for any \(t'\neq t\), \(\rOrder_t\) is independent of \(\rOrder_{t'}\). Furthermore, for any \(G' \in V\), \(\bP(\rOrder_t = G') = 1/m\).
\endproof
\proof[Proof of Lemma~\ref{lemma:glr-utility-lb-body}]
In this proof, we assume that the action profile selected \(\a\) is the group request strategy, hence, the set of valid groups is \(\G\). Fix an arbitrary group \(G_i\). To generate the sequence \(\rOrder \in \mathcal{S}_\G\) we use the algorithm from Proposition~\ref{prop:glr-random-order}, that is, generate a sequence \(\rOrder^{-}\) where \(\rOrder^{-}_t\) is independently and uniformly sample from \(\G\setminus G_i\) and then extend it to \(\rOrder\). Let \(S_n = \sum_{t=1}^n |\rOrder^{-}_t|\). We let \(\tau = \tau(k-|G_i|+1, \rOrder^{-})\) be as defined in \eqref{eq:def:tau} where the size function is the cardinality of each valid group. Intuitively, \(\tau\) is the number of positions in \(\rOrder\) that ensures a payoff of 1 to \(G_i\) given \(\rOrder^{-}\). Note that if \(G_i\) is in the first \(\tau\) positions of \(\rOrder\), then the number of tickets awarded before it is processed is at most \(k-|G_i|\). On the other hand, if it is processed after \(\tau\) groups this number is at least \(k-|G_i|+1\). Therefore, we get
\begin{equation}\label{glr-cond-utility}
u_i(\GLR(\a))  = \E[\E[u_i(\GLR(\a))|\rOrder^{-}]] = \E\left[1 - \left(1-\frac{1}{m}\right)^\tau\right].
\end{equation}
Thus, to prove equation~\eqref{eq:glr-utility-lb} it suffices to show
\begin{equation}\label{glr-exp-tau}
\E\left[1 - \left(1-\frac{1}{m}\right)^\tau\right] \ge \frac{k}{n}(1 - \kappa) g(\alpha).
\end{equation}
We let \(m_j\) be the number of groups of size \(j\) in \(\G\setminus G_i\), more precisely,
\[m_j = \sum_{G \in \G\setminus G_i} \mathbf1\{|G| = j\}.\]
From this definition, it immediately follows that
\begin{align}
m -1 &= \sum_{j\ge1} m_j,\label{glr-mj-1}\\
\sum_{j\ge1} m_j j &= n - |G_i|.\label{glr-mj-2}
\end{align}
Define
\begin{equation}
\phi(\theta)=\E[e^{|\rOrder^{-}_1|\theta}] = \sum_{j\ge1} \left(\frac{m_j}{m-1}\right)e^{j\theta},
\end{equation}
We let \(F = \{F_n\}_{n\in\N}\) be the filtration generated by \(\rOrder^{-}\).
For any \(\theta \in \R\), we define the following martingale w.r.t. \(F_n\):
\begin{align}
\frac{e^{\theta S_n}}{\phi(\theta)^n}.
\end{align}
This expression is adapted with respect to \(F_n\), it is bounded as \(|\rOrder^-_i| \le \max_G|G|\) and, as shown below, it satisfies the martingale property:
\[\E\left[\frac{e^{\theta S_n}}{\phi(\theta)^n}|F_{n-1}\right] = \frac{e^{\theta S_{n-1}}}{\phi(\theta)^{n-1}}\E\left[\frac{e^{\theta |\rOrder^-_n|}}{\phi(\theta)}|F_{n-1}\right] = \frac{e^{\theta S_{n-1}}}{\phi(\theta)^{n-1}}\frac{\E\left[e^{\theta |\rOrder^-_n|}\right]}{\phi(\theta)}= \frac{e^{\theta S_{n-1}}}{\phi(\theta)^{n-1}}.\]
Clearly \(\tau\) is a stopping time w.r.t. \(F\), moreover, it is almost surely bounded because \(|\rOrder^-_i| \ge 1\) implies that \(\bP(\tau \le k-|G_i|+1) = 1\). Applying Doob's optional stopping theorem, we get
\begin{equation}\label{glr-doobs}
1 = \E\left[\frac{e^{\theta S_1}}{\phi(\theta)}\right] = \E\left[\frac{e^{\theta S_\tau}}{\phi(\theta)^\tau}\right].
\end{equation}
Moreover, if we restrict to \(\theta > 0\) and use that the definition of \(\tau\) implies
\[S_{\tau} \ge k-|G_i|+1,\]
we obtain
\begin{equation}\label{glr-stopped-mart}
\E\left[\frac{e^{\theta S_\tau}}{\phi(\theta)^\tau}\right]
\ge e^{\theta(k-|G_i|+1)}\E\left[\phi(\theta)^{-\tau}\right].
\end{equation}
Combining equations~\eqref{glr-doobs} and~\eqref{glr-stopped-mart} yields
\begin{equation}\label{glr-theta}
 e^{-\theta(k-|G_i|+1)}\ge \E\left[\phi(\theta)^{-\tau}\right].
\end{equation}
To prove equation~\eqref{glr-exp-tau}, we need an upper bound on \(\E\left[\left(1-\frac{1}{m}\right)^\tau\right]\). Thus, we let \(\theta^*\) be the unique solution of
\begin{equation}\label{glr-thetastar-def}
\phi(\theta) = \left(1 - \frac{1}{m}\right)^{-1} = \frac{m}{m-1}.
\end{equation}
The existence and uniqueness of \(\theta^*\) is guaranteed because \(\phi(\theta)\) is increasing and continuous, \(\phi(0) = 1\) and for \(\theta \ge0,~\phi(\theta) \ge e^{\theta}\) hence \(\phi(\log(\frac{m}{m-1})) \ge \frac{m}{m-1}\). Then equation~\eqref{glr-theta} evaluates to
\[e^{-\theta^*(k-|G_i|+1)} \ge \E\left[\phi(\theta^*)^{-\tau}\right] = \E\left[\left(1-\frac{1}{m}\right)^\tau\right].\]
This implies
\begin{equation}\label{glr-utility-theta-star}
\E\left[1 - \left(1-\frac{1}{m}\right)^\tau\right] \ge
1 - e^{-\theta^*(k-|G_i|+1)} = \theta^*(k-|G_i|+1)g(\theta^*(k-|G_i|+1)).
\end{equation}
The expression above is an increasing function of \(\theta^*\). Hence, if \(\theta^* \ge 1/n\) then~\eqref{glr-exp-tau} holds as
\begin{align}
\theta^*(k-|G_i|+1)g(\theta^*(k-|G_i|+1)) &\ge \frac{k-|G_i|+1}{n}g\left(\frac{k-|G_i|+1}{n}\right)\\
&\ge \frac{k - \max_G|G| + 1}{n} g\left(\frac{k-|G_i|+1}{n}\right),
\end{align}
% \na{The language above (and the period) makes it seem like we are done proving~\eqref{glr-exp-tau} in this case. In reality, the following sentence is part of that proof.}\cb{changed.}
and since \(g\) is a decreasing function we have
\begin{equation}
g\left(\frac{k-|G_i|+1}{n}\right) \ge g\left(\frac{k}{n}\right) \ge g(\alpha).
\end{equation}
Thus, we assume \(\theta^*<1/n\). Again, because \(g\) is a decreasing function it follows that
\begin{equation}
g(\theta^*(k-|G_i|+1)) \ge  g\left(\frac{k-|G_i|+1}{n}\right) \ge g\left(\frac{k}{n}\right) \ge g(\alpha).
\end{equation}
Therefore, it suffices to show that
\begin{equation}\label{glr-lb-nog}
\theta^*(k-|G_i|+1) \ge  \frac{k}{n}(1 - \kappa).
\end{equation}
From the definition of \(\theta^*\), we get
\begin{equation}\label{theta-m-relation}
\frac{m}{m-1} = \phi\left(\theta^*\right)
= \sum_{j\ge1}\frac{m_je^{j\theta^*}}{m-1}
\le \frac{1}{m-1}\sum_{j\ge1} \frac{m_j}{1-j\theta^*}.
\end{equation}
In the inequality we use that for any \(x < 1,~ e^x \le 1/(1-x)\). Observe that
\[j\theta^* < j/n \le \max_G|G|/n < 1.\]
The first inequality follows by our assumption \(\theta^* < 1/n\), the second an third as \[j \le \max_G|G| \le k < n.\] If we multiple both sides of~\eqref{theta-m-relation} by \((m-1)\) and subtract \((m-1)\) we obtain
\[1 \le \sum_{j\ge1} \frac{m_j}{1-j\theta^*} - (m-1) =  \sum_{j\ge1} \frac{m_j}{1-j\theta^*} -  \sum_{j\ge1} m_j = \sum_{j\ge1}\frac{m_jj\theta^*}{1-j\theta^*}.\]
The first equality follows from equation~\eqref{glr-mj-1}. Besides,
\[\sum_{j\ge1}\frac{m_jj\theta^*}{1-j\theta^*}\le \sum_{j\ge1}\frac{m_jj\theta^*}{1-\max_G|G|\theta^*} =  \frac{(n-|G_i|)\theta^*}{1-\max_G|G|\theta^*}.\]
In the first inequality we use that \(j \le \max_G|G|\). The equality follows from equation~\eqref{glr-mj-2}. Combining both expressions above yields
\begin{equation}\label{theta-star-raw-lb}
1-\max_G|G|\theta^* \le (n-|G_i|)\theta^*.
\end{equation}
Rearranging, we have
\begin{equation}
n\theta^* \ge 1-(\max_G|G| - |G_i|)\theta^* > 1-(\max_G|G| - |G_i|)/n,
\end{equation}
where the second inequality follows by the assumption \(\theta^* < 1/n\). Substituting this last inequality into the left hand side of~\eqref{glr-lb-nog}, we have
\begin{align}
\theta^*(k-|G_i|+1) &\ge \frac{k}{n}\left(1 - \frac{|G_i| - 1}{k}\right)\left(1 - \frac{\max_G|G| - |G_i|}{n}\right)\\
&\ge \frac{k}{n}\left(1 - \frac{|G_i| - 1}{k} - \frac{\max_G|G| - |G_i|}{n}\right).
\end{align}
The expression at the right hand side is decreasing in \(|G_i|\), hence, minimized at \(|G_i| = \max_G|G|\). Substituting \(|G_i| = \max_G|G|\) above yields
\begin{equation}
\frac{k}{n}\left(1 - \frac{\max_G|G|-1}{k}\right) \ge \frac{k}{n}\left(1 - \kappa\right).
\end{equation}
The inequality follows as our instance is in \(I(\kappa,\alpha)\), hence
\begin{equation}%\label{eq:1-kappa}
\frac{\max_G|G|-1}{k} \le \kappa.
\end{equation}
\endproof
\subsubsection{Tightness}
\begin{proposition}\label{prop:spl-perfomance-example}
For any \(\alpha, \kappa \in (0,1)\) and \(\epsilon>0\), there exists an instance in \(I(\kappa, \alpha)\) such that the utilization of the group request equilibrium outcome of the {\NameProposedMechanism} is less than \(g(\alpha) + \epsilon\).
\end{proposition}
\proof[Proof of Proposition~\ref{prop:spl-perfomance-example}]
Fix \(\alpha, \kappa \in (0,1)\) and \(\epsilon > 0\). For any instance \(I\), we let \(U(I)\) be the utilization of the group request equilibrium outcome under the \NameProposedMechanism. We will construct a sequence of instances \(\{I_\eta\}\) such that for any \(\eta\in\N\),
\[I_\eta \in I(\kappa,\alpha) \text{ and } \lim_{\eta\to\infty}U(I_\eta)\to g(\alpha).\] In \(I_\eta\), there are \(n_\eta = m_\eta s_\eta\) agents divided in \(m_\eta\) groups of size \(s_\eta\), and \(k_\eta = \alpha m_\eta s_\eta\) tickets. We define \(\{m_\eta\}, \{s_\eta\}\) to be increasing sequences of natural numbers that satisfy three conditions:
\begin{enumerate}
    \item Each instance has an integer number of tickets, that is, \(\{m_\eta\}\) must be such that
    \begin{equation} \alpha m_\eta \in \N.\label{alpha-cond1} \end{equation}
    \item Each instance is in \(I(\kappa,\alpha)\), i.e.,
    \begin{align*}
    \frac{k_\eta}{n_\eta} &\le \alpha,\quad
    \frac{s_\eta-1}{k_\eta} \le  \kappa.
    \end{align*}
    The first condition holds immediately as
    \[\frac{k_\eta}{n_\eta} = \frac{\alpha m_\eta s_\eta}{m_\eta s_\eta} = \alpha.\]
    To ensure the second condition, we will define \(m_1\) to be such that
    \begin{equation} \alpha m_1 \ge \kappa^{-1}.\label{alpha-cond2} \end{equation}
    Observe that
    \begin{align*}
    \frac{s_\eta-1}{k_\eta} &= \frac{s_\eta-1}{\alpha m_\eta s_\eta} \le \frac{1}{\alpha m_\eta} \le \frac{1}{\alpha m_1} \le \kappa.
    \end{align*}
    The second inequality follows as \(\{m_\eta\}\) is increasing, and the third by condition~\eqref{alpha-cond2}.
    \item Both sequences grow at a similar rate, more precisely, there exists a positive constant \(c\) such that
    \begin{equation} \frac{m_\eta}{s_\eta} \le c. \label{seqs-cond3} \end{equation}
\end{enumerate}
We will define both sequences explicitly when \(\alpha\) is rational. In this case, there exists \(p,q\in\N\) such that \(\alpha= p/q\). Then, we can let
\begin{align*}
m_\eta = 2\eta q \lceil \kappa^{-1}\rceil,\quad
s_\eta = \eta q \lceil \kappa^{-1}\rceil.
\end{align*}
It's easy to see that conditions~\eqref{alpha-cond1},~\eqref{alpha-cond2} and~\eqref{seqs-cond3} holds, in the third condition \(c=2\). If \(\alpha\) is irrational, then we can choose a rational number \(\alpha^*\le \alpha\) that is arbitrarily close to \(\alpha\). We let the number of tickets be \(k_\eta = \alpha^* m_\eta s_\eta\) and define \(m_\eta\) in the same way as before but with respect to \(\alpha^*\).\\

Under the \NameProposedMechanism, we will draw \(\alpha m_\eta\) agents that get a full allocation. In this context, the utility of each agent is
\begin{equation}
u_i(\SPL_i(\a)) = 1 - \prod_{i=0}^{\alpha m_\eta-1}\left(1-\frac{1}{m_\eta-i/s_\eta}\right).
\end{equation}
Therefore, the utilization of this system correspond to
\begin{equation}
U(I_\eta) = \frac{n_\eta}{k_\eta}\left(1 - \prod_{i=0}^{\alpha m_\eta-1}\left(1-\frac{1}{m_\eta-i/s_\eta}\right)\right) = \frac{1}{\alpha}\left(1 - \prod_{i=0}^{\alpha m_\eta-1}\left(1-\frac{1}{m_\eta-i/s_\eta}\right)\right).
\end{equation}
We claim that
\[\lim_{\eta\to\infty}U(I_\eta) \to g(\alpha).\]
Observe that
\begin{align*}
\lim\inf_{\eta\to\infty}\prod_{i=0}^{\alpha m_\eta-1}\left(1-\frac{1}{m_\eta-i/s_\eta}\right)
&\ge \lim_{\eta\to\infty}\left(1-\frac{1}{m_\eta}\right)^{\alpha m_\eta} = e^{-\alpha},\\
\lim\sup_{\eta\to\infty}\prod_{i=0}^{\alpha m_\eta-1}\left(1-\frac{1}{m_\eta-i/s_\eta}\right)
&\le \lim_{\eta\to\infty}\left(1-\frac{1}{m_\eta-\frac{\alpha m_\eta}{s_\eta}  }\right)^{\alpha m_\eta-1}\\
&\le \lim_{\eta\to\infty}\left(1-\frac{1}{m_\eta-\alpha c}\right)^{\alpha m_\eta-1}
 = e^{-\alpha}.
\end{align*}
The last inequality follows by condition~\eqref{seqs-cond3}.
\endproof

\end{document}